\documentclass[preprint,12pt]{elsarticle}
\usepackage{graphicx}
\usepackage{amssymb}
\usepackage{amsmath}
\usepackage{subfig}
\usepackage{multirow}
\usepackage{colortbl,color}
\usepackage{algorithm}
\usepackage[page,title,titletoc,header]{appendix}
\usepackage{multirow}
\usepackage{amsthm}
\usepackage{hyperref}
\usepackage{bm}
\biboptions{comma,sort&compress}
\journal{ }

\begin{document}

\begin{frontmatter}
\title{Numerical study on thermal transpiration flows through a rectangular channel}
\author[KFUPM]{Jun~Li}
\author[MTU]{Chunpei~Cai}
\author[CARDC]{Zhi-Hui Li}
\address[KFUPM]{Center for Integrative Petroleum Research, \\ College of Petroleum Engineering and Geosciences, \\ King Fahd University of Petroleum $\&$ Minerals, Saudi Arabia}
\address[MTU]{Department of Mechanical Engineering-Engineering Mechanics, \\ Michigan Technological University, Houghton, MI 49931, USA}
\address[CARDC]{Hypervelocity Aerodynamics Institute, \\ China Aerodynamics Research and Development Center, \\ Mianyang 621000, China}
\begin{abstract}
Gaseous thermal transpiration flows through a rectangular micro-channel are simulated by the direct simulation BGK (DSBGK) method. These flows are rarefied, within the slip and transitional flow regimes, which are beyond many traditional computational fluid dynamic simulation schemes, such as those based on the continuum flow assumption. The flows are very slow and thus many traditional particle simulation methods suffer large statistical noises. The adopted method is a combination of particle and gas kinetic methods and it can simulate micro-flows properly. The simulation results of mass flow rates have excellent agreement with experimental measurements. In another case of 2D channel, the DSBGK comparisons with the DSMC result and the solution of Shakhov equation are also in very good agreement. Another finding from this study is that numerical simulations by including two reservoirs at the channel ends lead to appreciable differences in simulation results of velocity and pressure distributions within the micro-channel. This is due to the inhaling and exhaling effects of reservoirs at the channel ends. Even though excluding those reservoirs may accelerate the simulations significantly by using a single channel in simulations, special attentions are needed because this treatment may over-simplify the problem, and some procedures and results may be questionable. One example is to determine the surface momentum accommodation coefficient by using analytical solution of the mass flow rate obtained in a single-channel problem without the confinement effect of reservoirs at the two ends.
\end{abstract}
\end{frontmatter}

\section{Introduction}\label{s:Introduction}
It is well known that rarefied gas flows through a tube with a constant pressure but variable temperature along the wall boundary may experience appreciable bulk speed \cite{Reynolds,Maxwell,Knudsen}. In fact, small scale gas flows within micro-channels, including micro- thermal transpiration flows, may have high rarefaction effect, which is a challenge for investigations. These effects can be characterized by the Knudsen number ($Kn$) \cite{Shen2005, Ali}:
\begin{equation}
      Kn = \lambda /L,
\end{equation}
where $\lambda$ is the molecular mean free path of gas, and $L$ is a characteristic length, which can be the micro-channel height. According to different $Kn$ numbers, gas flows can be continuum  ($Kn<0.01$), slip ($0.01 < Kn <0.1$), transitional ($0.1<Kn <10$), and  collisionless ($Kn>10$). Micro and thermal transpiration flows can be within any of these regimes with $Kn>0.01$.

The Micro- /Nano- Electro-Mechanical-Systems (MEMS/NEMS) have decreased to sub-microns in recent decades, where $Kn$ can be large enough to be transitional. As such, thermal transpiration flows have more applications and become more important. For example, using the pumping effects of thermal transpiration to create micro-compressor without moving parts leads to the work of Vargo \cite{Vargo}, Young \cite{Young} and Alexeenko \cite{Alexeenko}. Gupta and Gianchandani \cite{Gupta} developed a 48 multi-stage Knudsen compressor for on-chip vacuum resulting in compression ratios up to 50.   It is easy to understand that further investigations on thermal transpiration flow are necessary, and this is the major goal of this paper.

The rest of this paper is organized as follows. Section \ref{s:review} reviews related past work; Section \ref{s:DSBGK method} discusses the numerical method used in this study, i.e., the direct simulation BGK (DSBGK) method; Section \ref{s:Schematic} presents the simulation schematic used to mimic the experiments; Section \ref{s:Test smaller cases} shows the test cases with a small micro-channel to study the effects of reservoirs connected at the channel ends; Section \ref{s:Real cases} gives the simulation results of a real micro-channel with validation against experimental data.  The last section summarizes this study with several conclusions. 

\section{Related past work}\label{s:review}
In the literature, there are studies on gaseous flows at micro scale, including experimental measurements and numerical simulations.   These micro- flows can be pressure-driven flows and thermal transpiration flows.  Here we only name a few.

Experimental measurements are valuable to study micro-flows. Interesting phenomena are observed and can offer valuable physical insights and benchmarks to test simulations. In  many situations, experiential studies are not replaceable. Liang \cite{Liang} analyzed the behavior of the Thermal Pressure Difference (TPD) and the Thermal Pressure Ratio (TPR) for different gases by applying various temperature differences, searched a correction factor for pressure measurements, and obtained an easy-to-use equation than the Weber and Schmidt's ones \cite{Weber}. Later, Rosenberg and Martel \cite{Rosenberg} performed and compared their own measurements with those by Weber, Schmidt and Liang.  Marcos studied unsteady thermal transpiration rarefied gas flows inside a micro-tube and a micro-rectangular channel \cite{marcros1,marcros2,marcros3}. They found that the unsteady pressure developments in the two reservoirs at the micro-channel ends can be well approximated with two exponential functions.  They measured the slops of initial pressure changes inside the two reservoirs for flows with different degrees of rarefaction, the TPR, TPD and the thermal-molecular pressure ratio $\gamma$. Los and Fergusson \cite{los} noted the existence of a maximum value in their  TPD results. Takaishi and Sensui \cite{Takaishi} improved Liang's law. Annis \cite{annis} compared his measurements with the numerical results of Loyalka and Cipolla \cite{Loyalka}, and found his results are quite different from Maxwell's initial approach. Sone and Sugimoto \cite{sone} and Sugimoto \cite{Sugimoto} performed original experiments, using a micro-windmill set at the end of a bent capillary, allowing qualitative but not quantitative analysis of the mass flow rate induced by thermal transpiration. Variation work \cite{Ewart} derived from the constant volume technique tracked pressure variation with time at the inlet and outlet of the tube, and obtained the mass flow rate, which is related to the pressure variations.  The most recent experimental work on thermal transpiration probably is the measurement of the mass flow rates of gas flows through a rectangular channel by measuring the initial pressure change rate inside the reservoirs at the channel ends \cite{Yamaguchi2014, Yamaguchi2016}.

There are many Computational Fluid Dynamics (CFD) schemes, which can be categorized into three classes. The first class is on the macroscopic level and applicable to simulating flows in the continuum and slip regimes, where the governing equations is usually the Navier-Stokes equation or the Burnett equation, and the non-slip boundary condition or general velocity-slip and temperature-jump boundary conditions shall be used. But, it is improper to use these CFD schemes in transitional and collisionless flow regimes, which may happen in thermal transpiration flows. The second class is on the mesoscopic level and based on the gas kinetic theory and velocity distribution functions. The fundamental governing equation is the Boltzmann equation or its simplified Bhatnagar-Gross-Krook (BGK) model \cite{BGK}. Related methods include the Lattice Boltzmann Method \cite{Qian1992,luo,chen}. Graur and Shripov \cite{Sharipov} used gas kinetic method to simulate rarefied gas flows along a long pipe with an elliptical cross-sections. There are also various kinds of hybrid methods that are based on the gas kinetic theory, such as the so-called gas kinetic scheme (GKS) or unified gas kinetic scheme (UGKS) \cite{xugks,xu,xuuks},  which are applicable to the simulations of micro-flows. The GKS method continues to re-construct the velocity distribution function at the mesoscopic level, based on which the mass, momentum and energy fluxes can be computed correctly, and then the macroscopic properties are updated, such as density, velocity, pressure and temperature.  The last class of CFD schemes is based on the molecular dynamics, which targets each molecule or atom. One example is the direct simulation Monte Carlo (DSMC) method \cite{Bird1963,Bird1994}.

This paper aims to report investigations on thermal transpiration flows with a specific numerical simulation method to be discussed in the next section. Gaseous micro-flows usually have different degrees of rarefaction and thus cannot be properly modeled by the first class of CFD schemes. The third type of methods is demanding because it traces particles' movements and computes particles' collisions in a statistical approach. However, the common issue associated with the traditional particle methods is the large statistical noises in low-speed gas flows including microchannel flows, and much effort has been spent to reduce these noises, such as the Information Preservation method \cite{fan0,cai,fan1,sun,fan2}. The second class of methods solves the velocity distribution function and is also quite demanding, especially when the intermolecular collisions are considered.

\section{The DSBGK method}\label{s:DSBGK method}
The direct simulation BGK (DSBGK) method was proposed recently \cite{Li2010, Li2012}, and it is based on the BGK model for the Boltzmann equation. The BGK model approximates the standard Boltzmann equation quite well in rarefied gas problems with small perturbations, where the solution of distribution function is close to the local Maxwell velocity distribution.

The thermal transpiration phenomenon is simulated here at different pressure conditions and for different gas species by the Fortran MPI software package \textit{NanoGasSim} developed using the DSBGK method. As a molecular simulation method, the DSBGK method works like the standard DSMC method \cite{Bird1963,Bird1994} but actually is a rigorous mathematical solver of BGK-like equation, instead of physical modeling of the molecular movements. At the initial state, the computational domain is divided uniformly in each direction into many cells, which are either void and solid. About twenty simulated molecules are randomly distributed inside each void cell and assigned with initial positions, velocities and other molecular variables according to a specified initial probability distribution function. The cell size and time step are selected the same as in the DSMC simulations. During each time step, each simulated molecule moves uniformly and in a straight line before randomly reflecting at the wall surface and its molecular variables are updated along each segment of the trajectory located inside a particular void cell according to the BGK equation. At the end of each time step, the number density, flow velocity and temperature at each void cell are updated using the increments of molecular variables along these segments located inside the concerned cell according to the conservation laws of mass, momentum and energy of the intermolecular collision process.

The major differences of the DSBGK method from the traditional DSMC method are: 1) the DSMC method uses the transient values of molecular variables to compute the cell's variables, which are subject to large stochastic noise due to random and frequent molecular movements into and out of each cell, while the DSBGK method employs the \textit{increments} of molecular variables due to intermolecular collisions to \textit{update} the cell's variables based on the conservations laws mentioned above; 2) the DSBGK method computes the effect of intermolecular collisions by solving the BGK model while the DSMC method randomly handles the intermolecular collision effect using an importance sampling scheme, which costs a noticeable percent of computational time to generate a huge number of random fractions. These two differences significantly improve the efficiency of DSBGK method particularly at low speed conditions such as the micro thermal transpiration flows.

More algorithm discussions as well as the convergence proof of the DSBGK method are detailed in \cite{Li2012}. The DSBGK method has been comprehensively verified against the DSMC method over a wide range of $Kn$ number in several benchmark problems and is much more efficient than the DSMC method. Recently, the DSBGK method was successfully applied to study shale gas flows inside a real three-dimensional digital rock sample with 100-cubed voxels over a wide range of $Kn$ \cite{LiSultan2015} and the Klinkenberg slippage effect in the computation of apparent permeability \cite{LiSultan2016}.

It is also important to emphasize the differences between LBM and DSBGK method. The former can have superior parallel computing performance, however, it suffers a severe issue in discretizing the molecular velocity space due to its simplicity in algorithm. The velocity space discretization in LBM is rather simple and thus it is crude. LBM sacrifices the physical accuracy to achieve the mathematical simplicity. For example, in two-dimensional flows, the ordinary LBM adoptes the D2Q9 model, where only 9 points are used to discretize the whole velocity distribution function \cite{Qian1992}. By contrast, the DSBGK method uses dynamic molecular velocities to discretize the velocity space, which is physically more accurate. It allows as fine discretization as desired since the molecular velocity set used in the discretization is dynamically updated during the simulation as in the DSMC simulation.

\section{Simulation schematic}\label{s:Schematic}
\begin{figure}[H]
	\centering
	\includegraphics[width=0.52\textwidth]{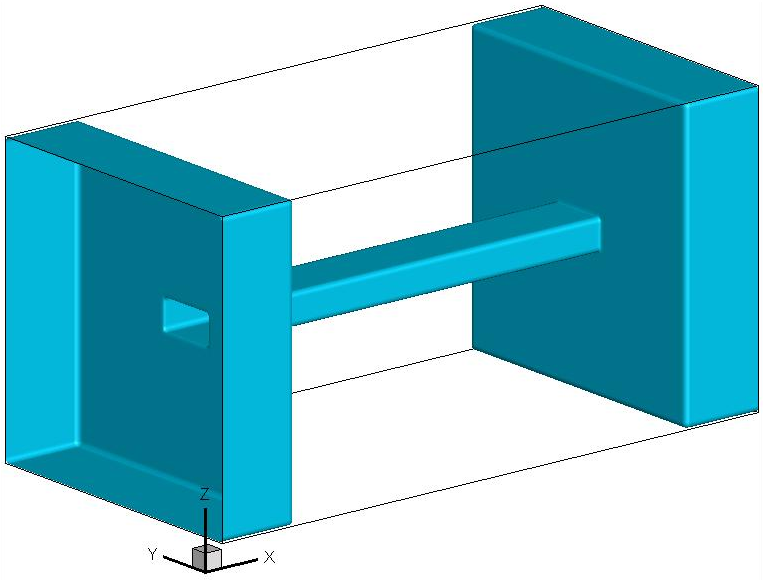}
	\includegraphics[width=0.42\textwidth]{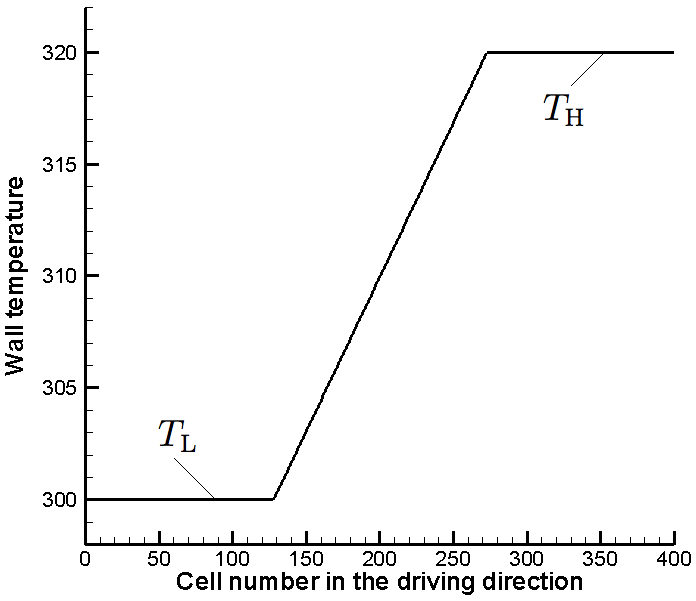}
	\caption{Micro-channel connected with reservoirs (left) and representative temperature distribution on the wall boundary (right).}
	\label{geometry and temperature}
\end{figure}

Fig. ~\ref{geometry and temperature} illustrates the simulation domain, which is similar to the configuration used in the gas flow experiments \cite{Yamaguchi2014, Yamaguchi2016}. Two reservoirs are connected to a micro- rectangular channel at the two channel ends.   Compared with the true dimensions adopted in experiments, the reservoir sizes are decreased to reduce the simulation cost. The sizes of these two reservoirs are chosen sufficiently large to make negligible the influence of reservoir sizes on the simulation results. These results include the pressure difference between the two reservoirs and the mass flow rate through the micro-channel. 

In these simulations, the wall temperature $T_{\rm wall}$ solely depends on the $x$ coordinate  and increases linearly from $T_{\rm L}$ to $T_{\rm H}$, which are the constant wall temperatures of the two reservoirs. This temperature difference drives the gas flow through the channel with the thermal transpiration effect.

The total length, width and height of the computational domain are denoted as  $L_{\rm all}$, $W_{\rm all}$ and $H_{\rm all}$, respectively.  $L_{\rm micro}$, $W_{\rm micro}$ and $H_{\rm micro}$ are three dimensions for the micro-channel. In the experiments, a micro-valve is used to close or open the passage between the two reservoirs. Correspondingly, the two boundaries at $x=0$ and $x=L_{\rm all}$ switch between wall boundaries and periodic boundaries in the simulations. When the micro-valve is close/open, the pressure difference/mass flow rate at steady state through the micro-channel depends on the initial Knudsen number $Kn_0$ and the temperatures of two reservoirs.

The pressure difference is studied by adopting a wall boundary at $x=L_{\rm all}$ and an open boundary at $x=0$. At $x=0$, the pressure is fixed as $p\equiv p_0$,  the temperature is set as $T\equiv T_{\rm wall}(x=0)$ and the transient flow velocity $\vec u=(u,v,w)$ is computed at the cell adjacent to $x=0$ the same as in the DSMC simulations.

To ease the numerical study, separate simulations are performed to compute the mass flow rates by using two open boundaries at $x=0$ and $x=L_{\rm all}$, respectively. The experiments \cite{Yamaguchi2014, Yamaguchi2016} can be conveniently modified to achieve the current simulation setups with at least one open boundary. The setups are close to those in real applications. It is not surprising to observe experimental and numerical results maybe not comparable,  which usually is due to different settings.  These discrepancies do not appear in the current simulations for the \textit{steady state} pressure difference after closing the micro-valve \cite{Yamaguchi2014} and the mass flow rate before closing the micro-valve \cite{Yamaguchi2016}.

\section{Simulation tests with smaller channel and temperature difference}\label{s:Test smaller cases}
Several basic parameters are listed here: the lowest temperature $T_{\rm L}$=300 K, the highest temperature $T_{\rm H}$=320 K, initial temperature $T_0\equiv (T_{\rm L}+T_{\rm H})/2$, and initial pressure $p_0 = 50$ Pa. The gas is argon,  the dynamic viscosity is $\mu=2.117\times10^{-5}\times(T/273)^{0.81} {\rm Pa\cdot s}$ and molecular mass is $m=66.3\times10^{-27}$ kg \cite{Shen2005}. At $T_0$ and $p_0$, the dynamic viscosity is $\mu_0=2.346\times10^{-5}{\rm Pa\cdot s}$ and the mean free path is $\lambda_0=0.1522$ mm.

The simulation time step is set as $\Delta t=0.8\lambda_0/\sqrt{2k_{\rm B}T_0/m}=0.3389\times10^{-6}$ s, where $k_{\rm B}$ is the Boltzmann constant. Correspondingly, the cell sizes are set as $\Delta x=\Delta y=\Delta z=0.05$ mm $<\lambda_0$. The sizes of micro-channel located at the center of the computational domain are $L_{\rm micro}\times W_{\rm micro}\times H_{\rm micro}\equiv7.3$ mm $\times 1$ mm $\times 0.5$ mm and there are $146\times 20\times 10$ cells located inside the micro-channel with $Kn_0=\lambda_0/H_{\rm micro}=0.3044$. 

In the DSBGK simulations, the relaxation parameter $\upsilon$ of BGK model is computed using the transient $T$, $\mu(T)$ and number density $n$ by $\upsilon=2nk_{\rm B}T/(3\mu)$ for each cell as discussed in Section 4 of \cite{Li2012}. About 20 simulated molecules per cell are used unless stated otherwise. 

\subsection{Simulations of pressure difference}\label{ss:Test pressure}
In the simulation with a wall boundary at $x=L_{\rm all}$ and an open boundary at $x=0$, the steady state is almost static inside the reservoirs and the presences of reservoirs might have negligible influence on the steady state pressure difference between the two reservoirs. Thus, two simulations are performed for comparison: 

Case 1, a full domain simulation with $L_{\rm all}\times W_{\rm all}\times H_{\rm all}$=$10$ mm $\times 5$ mm $\times 5$ mm and the total cell number is $200\times 100\times 100$ with porosity $\phi=0.2846$; 

Case 2, a reduced domain simulation with $L_{\rm all}\times W_{\rm all}\times H_{\rm all}$=$L_{\rm micro}\times W_{\rm micro}\times H_{\rm micro}$, where the two reservoirs at the channel ends are removed. 

\begin{figure}[H]
	\centering
	\includegraphics[width=0.49\textwidth]{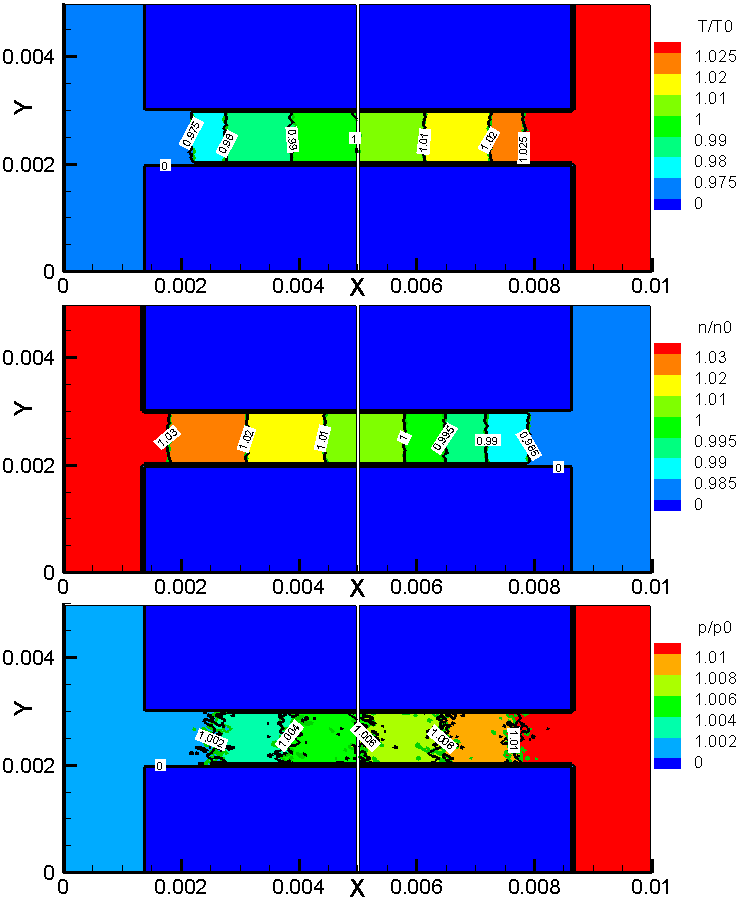}
	\includegraphics[width=0.49\textwidth]{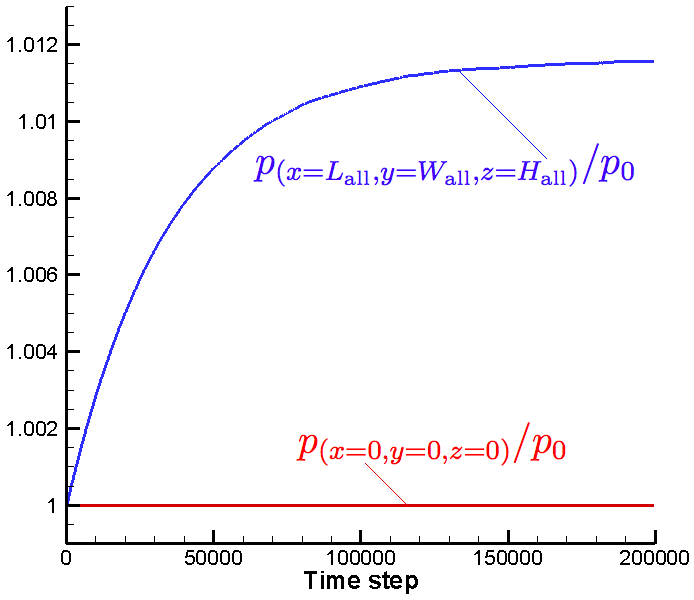}\\
	\caption{\textit{Transient} $T/T_0$, $n/n_0$ and $p/p_0$ on the middle XOY plane (left) , at $t=195,000 \Delta t$ (black lines) and $190,000 \Delta t$ (green lines); and the evolutions of $p/p_0$ at two corners of the simulation domain (right). Full domain simulation.} 
	\label{converge on XOY ib12TZ}
\end{figure}

Fig. \ref{converge on XOY ib12TZ} shows full domain simulation results. The left side displays the transient distributions of $T/T_0$, $n/n_0$ and $p/p_0$ at $t= 195,000 \Delta t$ and $190,000 \Delta t$ on the middle XOY plane with $z\equiv H_{\rm all}/2$. The results are represented with black and green lines, respectively, for different moments. As shown, at these two moments, the temperatures, number densities and pressures reached a steady state, which is consistent with the observation in the pressure evolutions with time at two points of the simulation domain as shown in Fig. \ref{converge on XOY ib12TZ} (right). The first point is at the bottom, front and left corner of the left reservoir,  and the second at the upper, back and right corner of the right reservoir. 

\begin{figure}[H]
	\centering
	\includegraphics[width=0.49\textwidth]{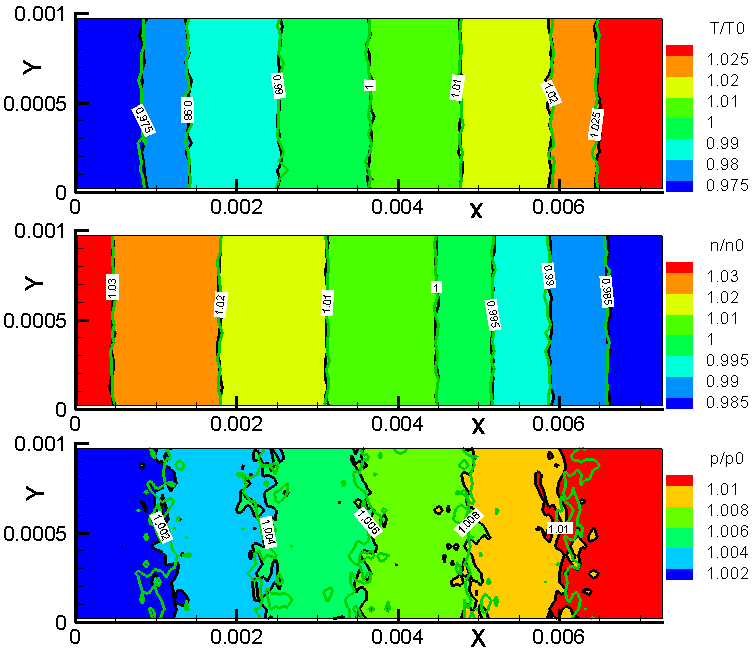}
	\includegraphics[width=0.49\textwidth]{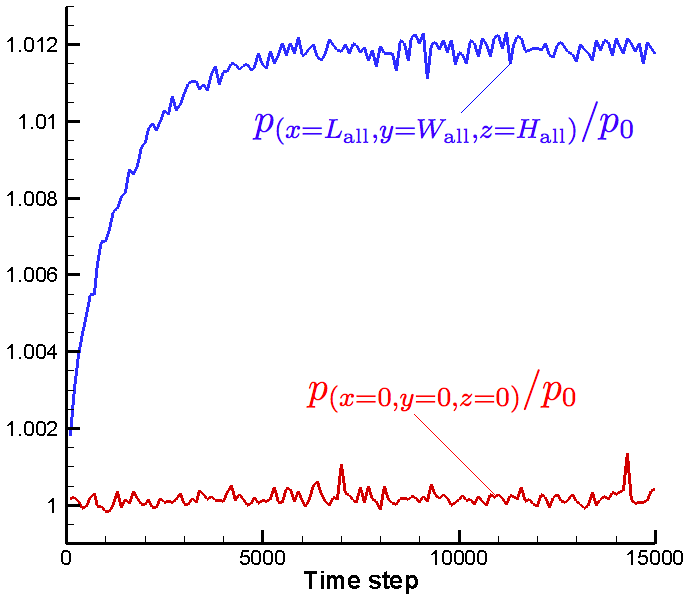}\\
	\caption{\textit{Transient} $T/T_0$, $n/n_0$, and $p/p_0$ on the middle XOY plane (left), at $t=10,000 \Delta t$ (black lines) and $15,000 \Delta t$ (green lines); and the evolutions of $p/p_0$ at two corners of the simulation domain (right). Reduced domain simulation.}
	\label{converge on XOY ib12TZnotanks}
\end{figure}

Fig. \ref{converge on XOY ib12TZnotanks} shows the results of reduced domain simulation. The left side displays the transient distributions of $T/T_0$, $n/n_0$ and $p/p_0$ on the middle XOY plane at $t= 10,000 \Delta t$ and $15,000 \Delta t$ using black and green lines, respectively. As shown, the temperatures, number densities and pressures reached a steady state after only $t=15,000 \Delta t$, which is also verified by the pressure evolutions shown in Fig. \ref{converge on XOY ib12TZnotanks} (right). The steady state inside the micro-channel is not static and thus the pressures collected at the ends of micro-channel contain obvious noises. To obtain smoother profiles, time average process is needed. 

The comparison between Figs. \ref{converge on XOY ib12TZ} and \ref{converge on XOY ib12TZnotanks} indicts that a full domain simulation needs much more time steps to converge, with a further increase of computational cost due to using a larger number of cells. Thus, the computational cost of full domain simulation is significantly higher than that of the reduced domain simulation. To save the simulation cost of studying the steady state pressure difference, a reduced domain simulation is favored and recommended.

\subsection{Simulation of mass flow rate}\label{ss:Test mass flow}
\begin{figure}[H]
	\centering
	\includegraphics[width=0.49\textwidth]{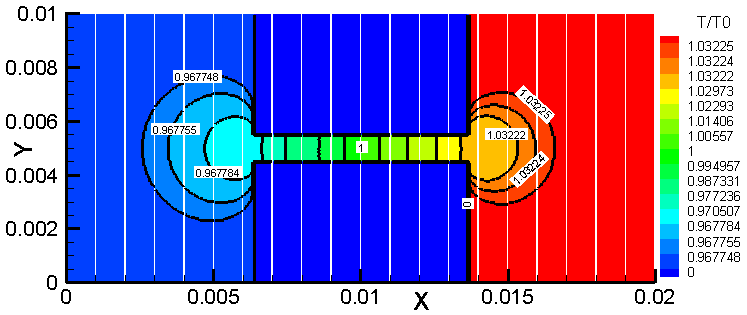}
	\includegraphics[width=0.49\textwidth]{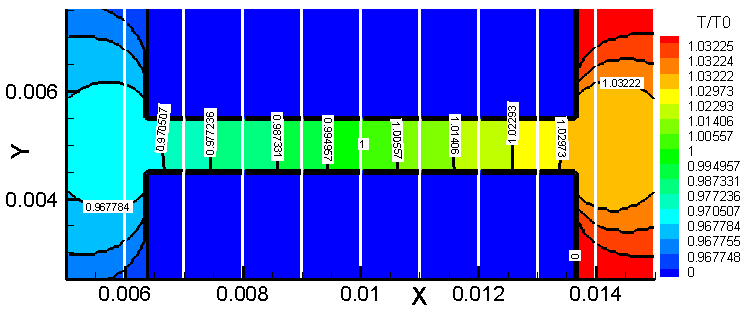}\\
	\includegraphics[width=0.49\textwidth]{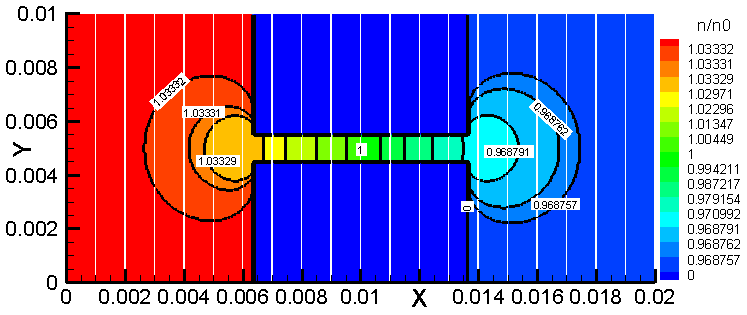}
	\includegraphics[width=0.49\textwidth]{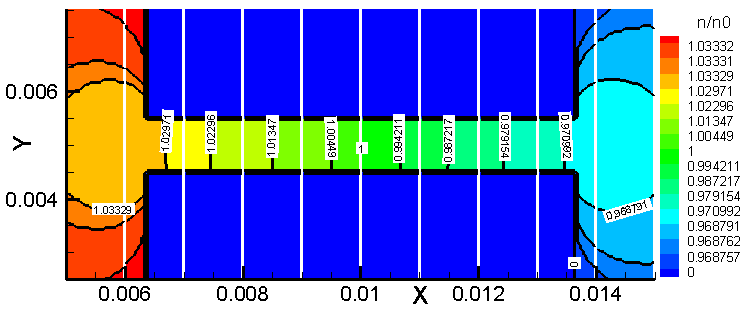}\\
	\includegraphics[width=0.49\textwidth]{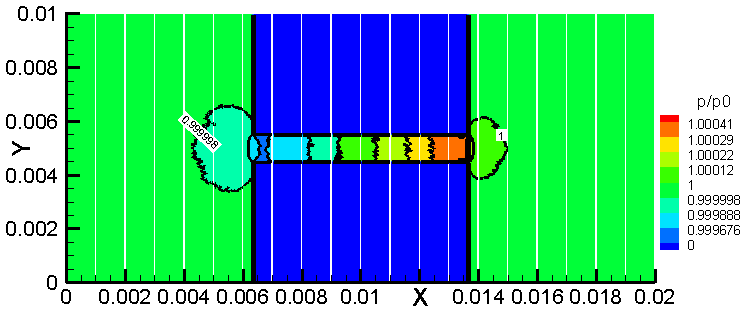}
	\includegraphics[width=0.49\textwidth]{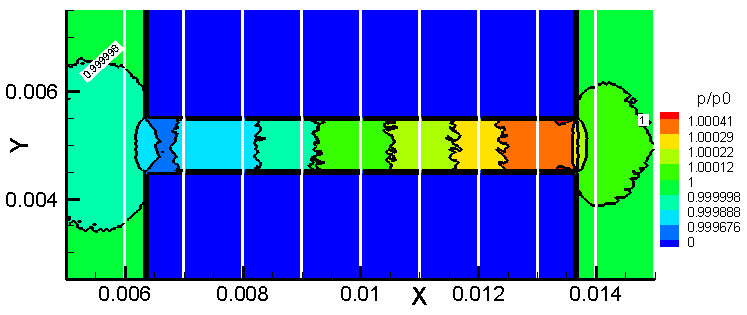}\\
	\includegraphics[width=0.49\textwidth]{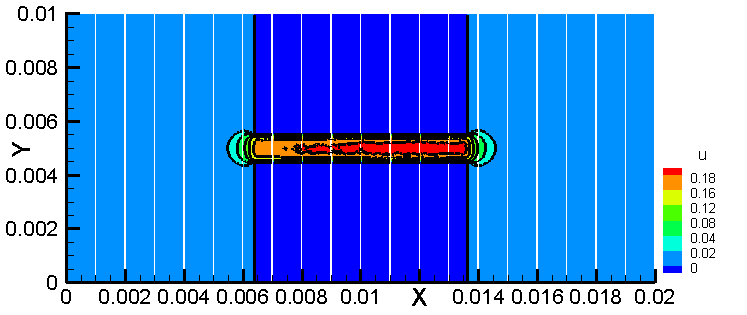}
	\includegraphics[width=0.49\textwidth]{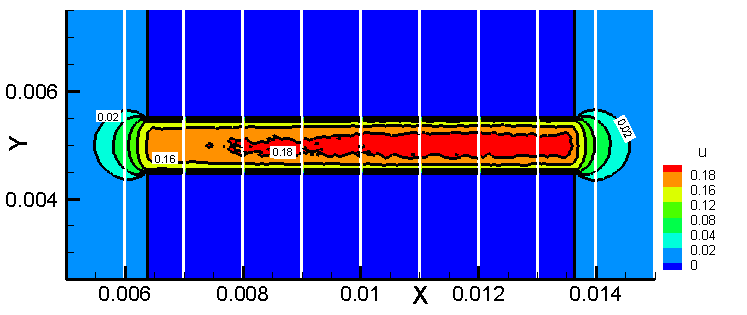}\\
	\includegraphics[width=0.49\textwidth]{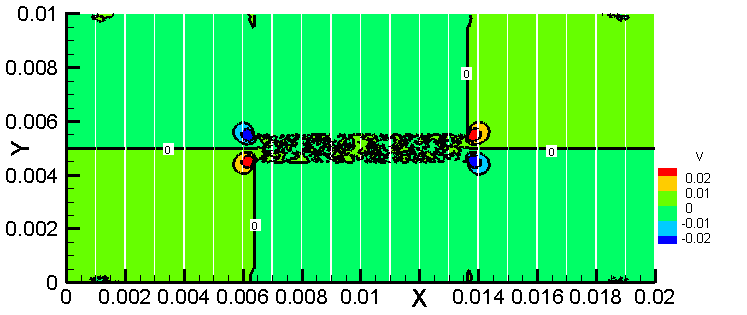}
	\includegraphics[width=0.49\textwidth]{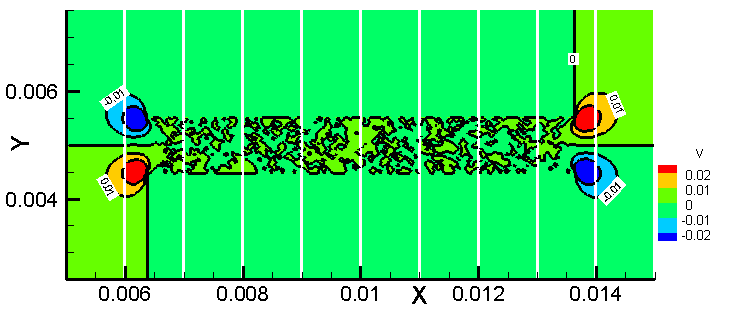}\\
	\caption{Average distributions of $T/T_0$, $n/n_0$, $p/p_0$, $u$ and $v$ on the middle XOY plane, using 5000 samples computed with $L_{\rm all}\times W_{\rm all}\times H_{\rm all}$=$20$ mm $\times 10$ mm $\times 5$ mm; left: distribution inside the whole plane, right: distribution inside a central subarea with fixed sizes of $10$ mm $\times 5$ mm.}
	\label{Average on XOY ib11TZlarge1}
\end{figure}

A full domain simulation with two open boundaries at $x=L_{\rm all}$ and $x=0$ is used to study the mass flow rate as mentioned before. To reduce the influence of reservoir sizes, we first use $L_{\rm all}\times W_{\rm all}\times H_{\rm all}$=$20$ mm $\times 10$ mm $\times 5$ mm. Time average process is used to smoothen the pressure and velocity distributions, as shown in Fig.~\ref{Average on XOY ib11TZlarge1}. Parallel computation is adopted because the total cell number and porosity are increased to $400\times200\times100$ and 0.63865, respectively. As shown by the white lines in Fig.~\ref{Average on XOY ib11TZlarge1}, the computational domain is divided only along the $x$ direction in the parallelization when visualization is needed.

\subsection{Influence of reservoir sizes on the mass flow rate}\label{ss:Size effect}
\begin{figure}[H]
	\centering
	\includegraphics[width=0.49\textwidth]{AverageXOYzoom-T-ib11TZlarge1.png}
	\includegraphics[width=0.49\textwidth]{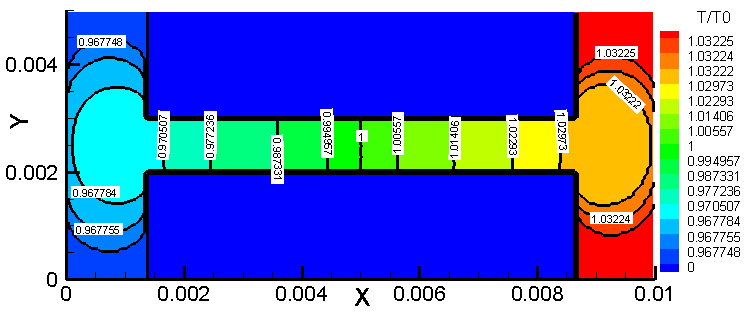}\\
	\includegraphics[width=0.49\textwidth]{AverageXOYzoom-n-ib11TZlarge1.png}
	\includegraphics[width=0.49\textwidth]{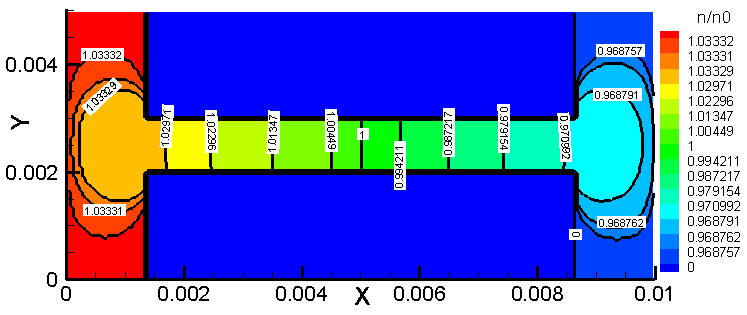}\\
	\includegraphics[width=0.49\textwidth]{AverageXOYzoom-p-ib11TZlarge1.png}
	\includegraphics[width=0.49\textwidth]{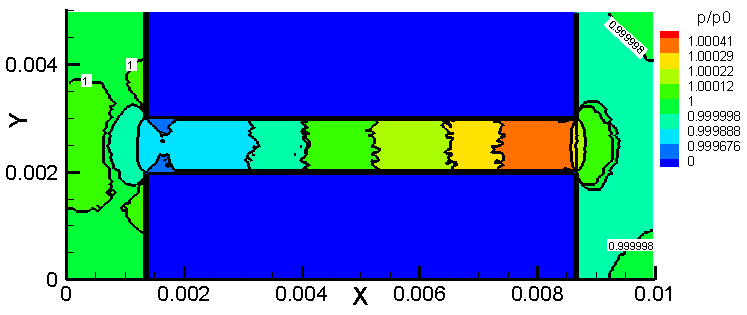}\\
	\includegraphics[width=0.49\textwidth]{AverageXOYzoom-u-ib11TZlarge1.png}
	\includegraphics[width=0.49\textwidth]{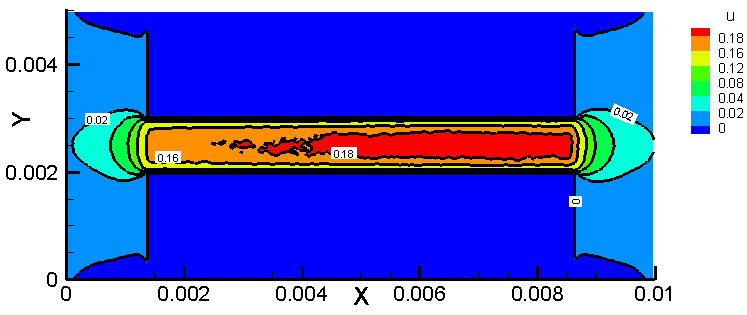}\\
	\includegraphics[width=0.49\textwidth]{AverageXOYzoom-v-ib11TZlarge1.png}
	\includegraphics[width=0.49\textwidth]{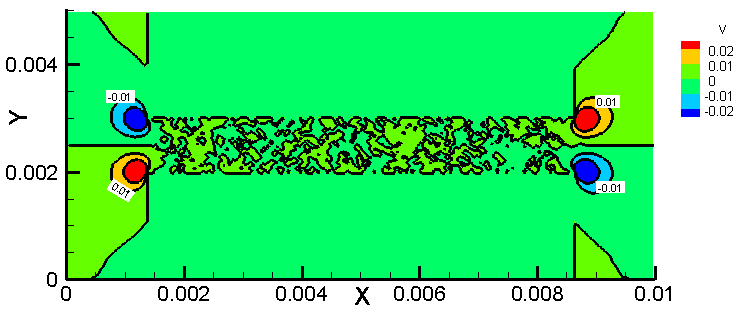}\\
	\caption{Average distributions of $T/T_0$, $n/n_0$, $p/p_0$, $u$ and $v$ on the middle XOY plane, using 5000 samples computed with $L_{\rm all}\times W_{\rm all}\times H_{\rm all}$=$20$ mm $\times 10$ mm $\times 5$ mm (left: a central subarea with fixed sizes of $10$ mm $\times 5$ mm) and $L_{\rm all}\times W_{\rm all}\times H_{\rm all}$=$10$ mm $\times 5$ mm $\times 5$ mm (right).}
	\label{Comparison between ib11TZlarge1 and ib11TZ on XOY}
\end{figure}

Fig.~\ref{Average on XOY ib11TZlarge1} shows that the reservoir sizes can be reduced to save computational cost. Accordingly, another full domain simulation with $L_{\rm all}\times W_{\rm all}\times H_{\rm all}$=$10$ mm $\times 5$ mm $\times 5$ mm is performed, and  Fig.~\ref{Comparison between ib11TZlarge1 and ib11TZ on XOY} shows the comparisons between the previous results of Fig.~\ref{Average on XOY ib11TZlarge1} (right) and the current results inside the same geometry configuration surrounding the micro-channel. 

Although the comparison shows that the solutions outside the micro-channel have appreciable differences, the solutions of $T/T_0$, $n/n_0$, $p/p_0$ and $u$ inside the micro-channel are almost the same when both computational domain sizes are not less than $L_{\rm all}\times W_{\rm all}\times H_{\rm all}$=$10$ mm $\times 5$ mm $\times 5$ mm (could be even smaller) for this particular micro-channel with $L_{\rm micro}\times W_{\rm micro}\times H_{\rm micro}$=$7.3$ mm $\times 1$ mm $\times 0.5$ mm. The magnitude of $v$ inside the micro-channel is too small to make comparison due to stochastic noise but the agreement of dominant $v$ outside the micro-channel is very good. Note that the micro-channel can be simplified by using periodic boundary conditions in the $y$ direction, when $W_{\rm micro}\gg H_{\rm micro}$ \cite{Yamaguchi2016}, where it only requires that the artificial reservoirs are much larger than the micro-channel in height. 

\subsection{Confinement effect of reservoirs on the mass flow rate}\label{ss:Confinement}
\begin{figure}[H]
	\centering
	\includegraphics[width=0.49\textwidth]{AverageXOY-T-ib11TZ}
	\includegraphics[width=0.49\textwidth]{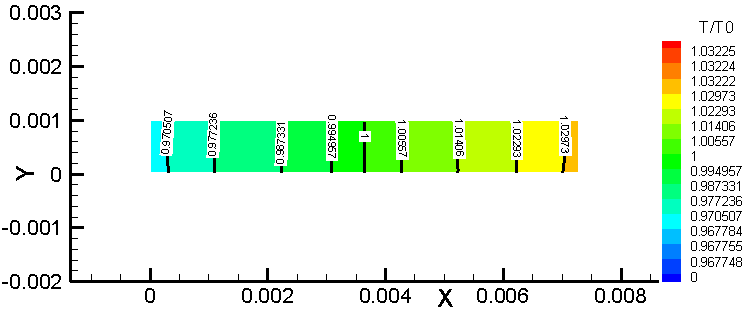}\\
	\includegraphics[width=0.49\textwidth]{AverageXOY-n-ib11TZ}
	\includegraphics[width=0.49\textwidth]{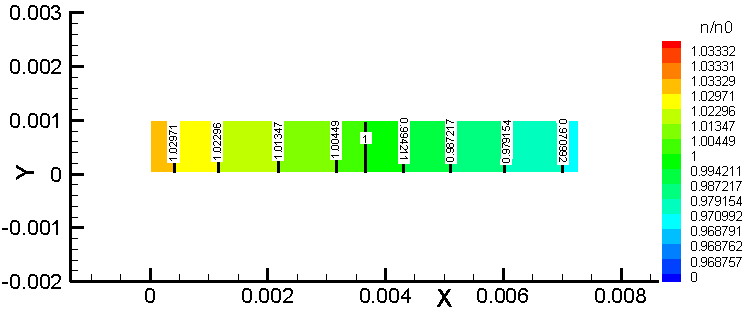}\\
	\includegraphics[width=0.49\textwidth]{AverageXOY-p-ib11TZ}
	\includegraphics[width=0.49\textwidth]{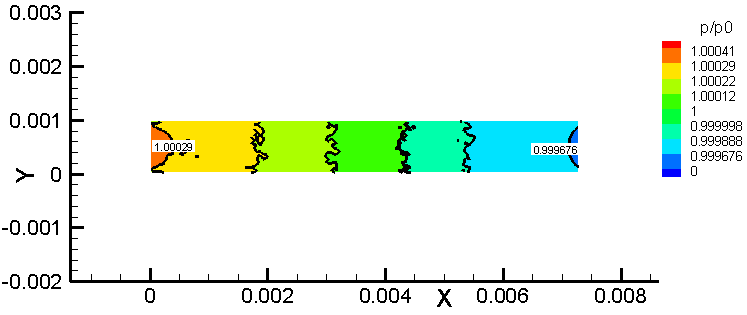}\\
	\includegraphics[width=0.49\textwidth]{AverageXOY-u-ib11TZ}
	\includegraphics[width=0.49\textwidth]{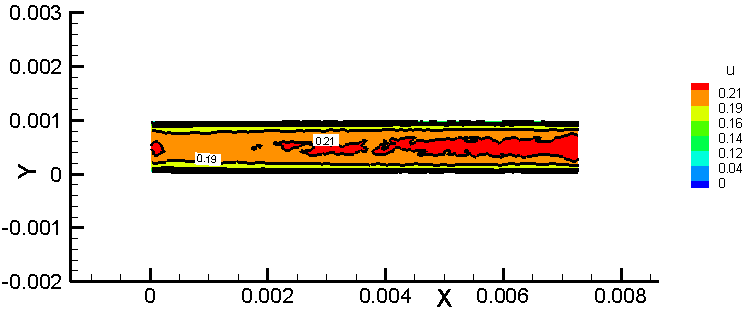}\\
	\includegraphics[width=0.49\textwidth]{AverageXOY-v-ib11TZ}
	\includegraphics[width=0.49\textwidth]{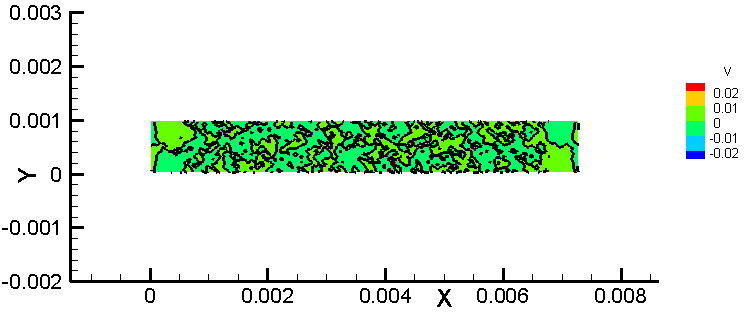}\\
	\caption{Average distributions of $T/T_0$, $n/n_0$, $p/p_0$, $u$ and $v$ on the middle XOY plane, using 5000 samples computed with $L_{\rm all}\times W_{\rm all}\times H_{\rm all}$=$10$ mm $\times 5$ mm $\times 5$ mm (left) and without reservoirs (right);  \textit{different} color legends are used in comparing $u$.}
	\label{Comparison between ib11 and notanks on XOY}
\end{figure}

\begin{figure}[H]
	\centering
	\includegraphics[width=0.45\textwidth]{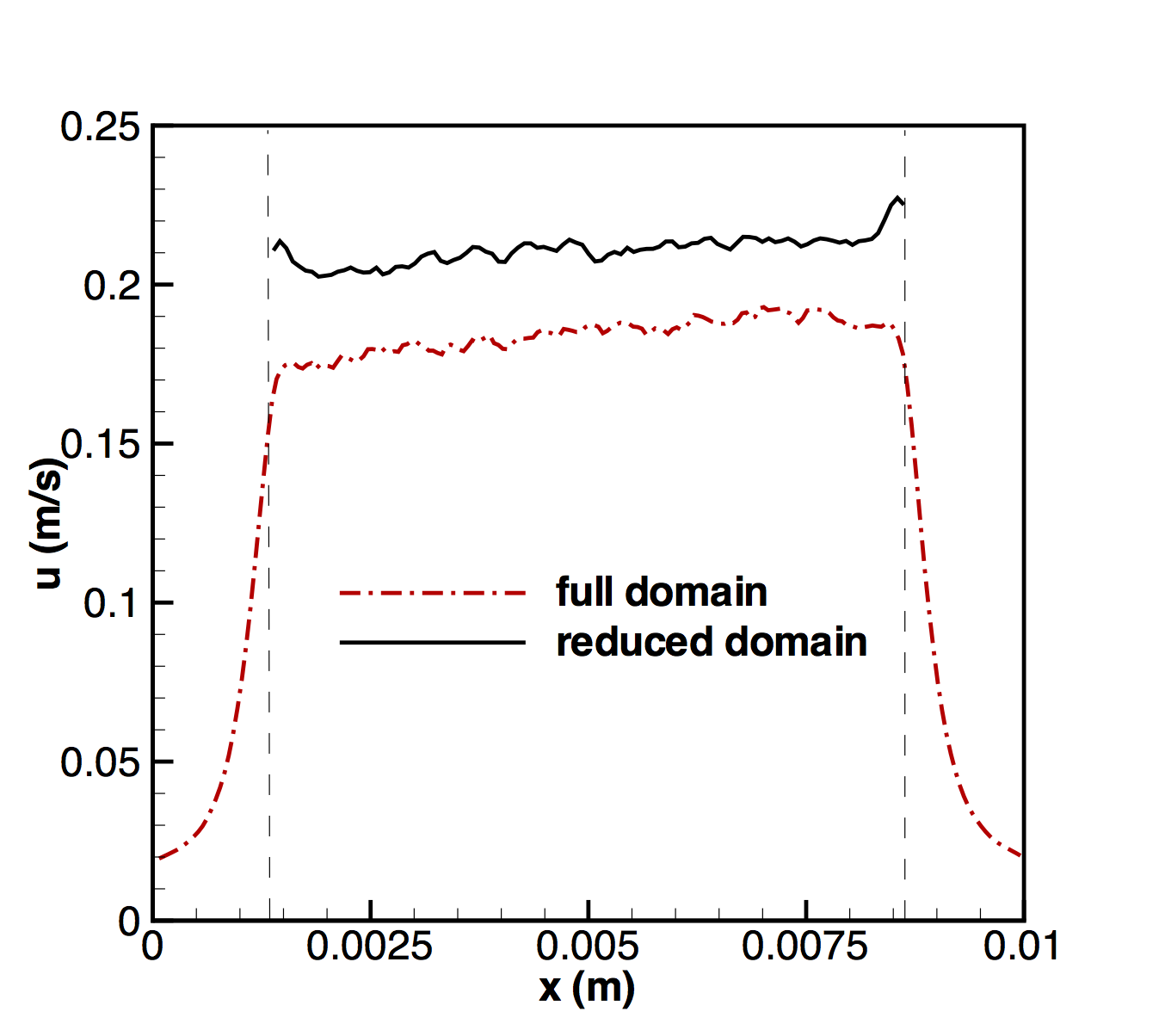}
	\includegraphics[width=0.45\textwidth]{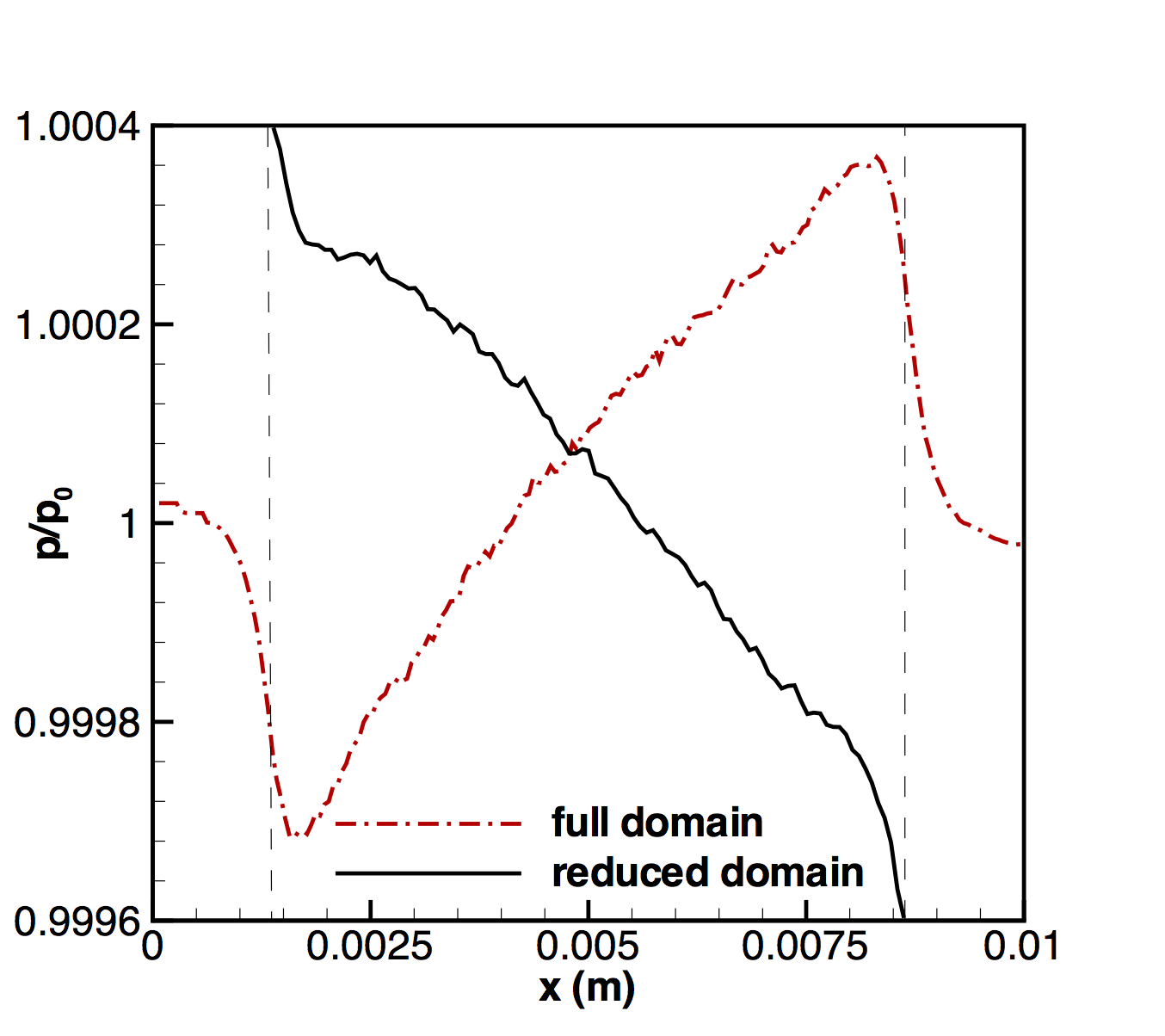}
	\caption{Velocity component $u$ (left) and pressure (right) along the micro-channel centerline with $y\equiv W_{\rm all}/2$ and $z\equiv H_{\rm all}/2$; vertical dash lines: locations of micro-channel ends.}
	\label{Fig:flatup}
\end{figure}

Usually there are entrance and exit effects at the two ends of micro-channel and the rarefaction effect further complicates the results \cite{entrance1,entrance2}. On the other hand, as shown in Section \ref{ss:Test pressure}, it also is desirable to accurately simulate the mass flow rate through the micro-channel without reservoirs, which make the simulation time-consuming. Thus, the micro-channel is modeled \textit{alone} with two open boundaries. The comparisons between the previous results computed with $L_{\rm all}\times W_{\rm all}\times H_{\rm all}$=$10$ mm $\times 5$ mm $\times 5$ mm (i.e., Fig.~\ref{Comparison between ib11TZlarge1 and ib11TZ on XOY} (right)) and the current results computed without reservoirs are given in Fig.~\ref{Comparison between ib11 and notanks on XOY}. The agreements between the two simulations for both $T/T_0$ and $n/n_0$ distributions inside the micro-channel are quite good. However, the current magnitude of dominant $u$ (about 0.19 m/s) is noticeably larger than the previous one (about 0.16 m/s) inside the micro-channel, which is consistent with the discrepancy in the comparison of $p/p_0$. According to Eq.~(3.4) of \cite{Yamaguchi2016}, the pressure gradient in the negative direction of $x$ axis enhances the mass flow rate in the current simulation, however, the pressure gradient in the positive direction of $x$ axis depresses the mass flow rate in the previous simulation, as shown in Fig.~\ref{Fig:flatup} (right). These two facts lead to a higher mass flow rate in the current simulation since the contribution of temperature gradient to the mass flow rate is almost the same. 

The presences of reservoirs require the inhaling and blowing effects near the two ends of the micro-channel to maintain the flow inside the two reservoirs with constant wall temperatures, which implies that the pressure gradient inside the micro-channel is \textit{certainly} in the positive direction of $x$ axis since the pressures at $x=0$ and $x=L_{\rm all}$ are equal to $p_0$. Note that the \textit{concentrated} pressure variations created by the inhaling and blowing effects depend mostly on the mass flow rate as long as the reservoirs are much larger than the cross-section of micro-channel, which interprets the good agreement of pressure differences across the micro-channel computed using different reservoir sizes as shown in Fig.~\ref{Comparison between ib11TZlarge1 and ib11TZ on XOY}. Thus, the confinement effect due to the presences of reservoirs as in the experiments \cite{Yamaguchi2016} always leads to a pressure gradient in the driving direction inside the micro-channel, which depresses the mass flow rate (e.g., the reduction could be as large as $\Delta u\approx (0.19-0.16)$ m/s for this particular case). This confinement effect should be reflected in the simulations by adding reservoirs into the configuration even though the objective is to study the flow quantities that depend mostly on the properties of gas (e.g., molecular species and pressure) and micro-channel (e.g., sizes and temperature distribution on the wall). This observation also implies that we need to be cautious when using the experimental data measured with the confinement effect to extrapolate the accommodation coefficients \cite{coeff} by using the analytical solution obtained from a single-channel problem without the confinement effect \cite{Yamaguchi2016}, unless it is intended to use these coefficients in the same analytical solution to predict the performances of similar micro-channels. As shown in Section~\ref{s:Real cases}, the results computed using the Maxwell diffuse reflection model (i.e., complete accommodation) agree well with the experimental data.

Fig.~\ref{Fig:flatup} shows velocity $u$ (left) and pressure (right) profiles along a micro-channel centerline extracted from Fig.~\ref{Comparison between ib11 and notanks on XOY}. The two vertical dash lines mark the channel entrance and exit. As shown, the average velocities over the micro-channel centerline have appreciable difference.  A quick estimation indicates that the peak values are 0.21 m/s and 0.18 m/s, respectively, or  a difference of 17\% when the full domain simulation is chosen as the reference because it is closer to the reality. In the full domain simulation, the velocity outside the micro-channel is small due to the large cross-section of the reservoirs. Fig.~\ref{Fig:flatup} (right) shows the pressure profiles along the micro-channel centerline. In this thermal transpiration flow, the pressure through the channel is almost constant, and the maximum relative variation of pressure is about 0.04 \% along the centerline.  Inside the micro-channel, even though the pressure variations are small, the difference between the two profiles is striking with completely opposite variation trends as discussed above.

\begin{figure}[H]
	\centering
	\includegraphics[width=0.45\textwidth]{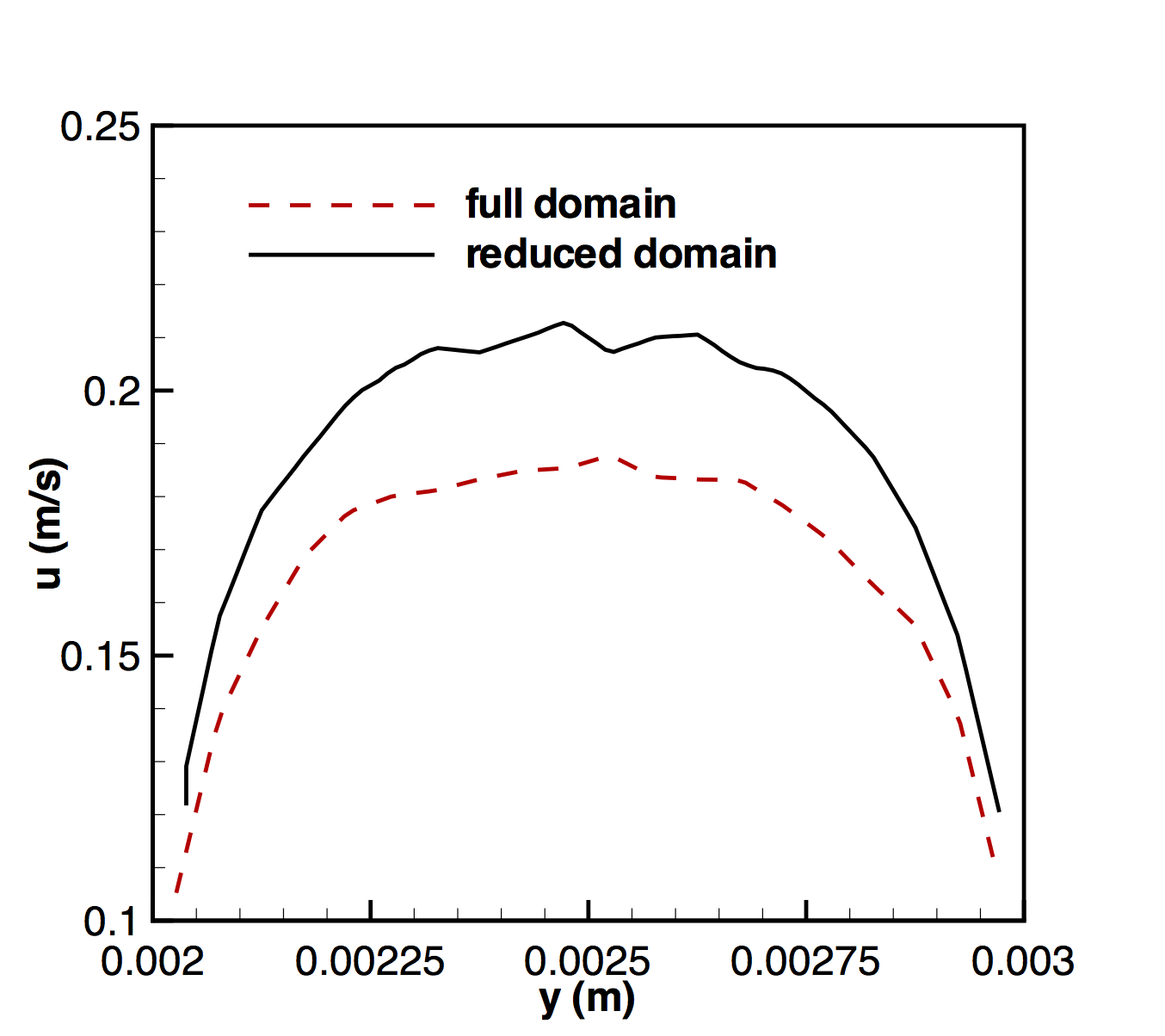}
	\includegraphics[width=0.45\textwidth]{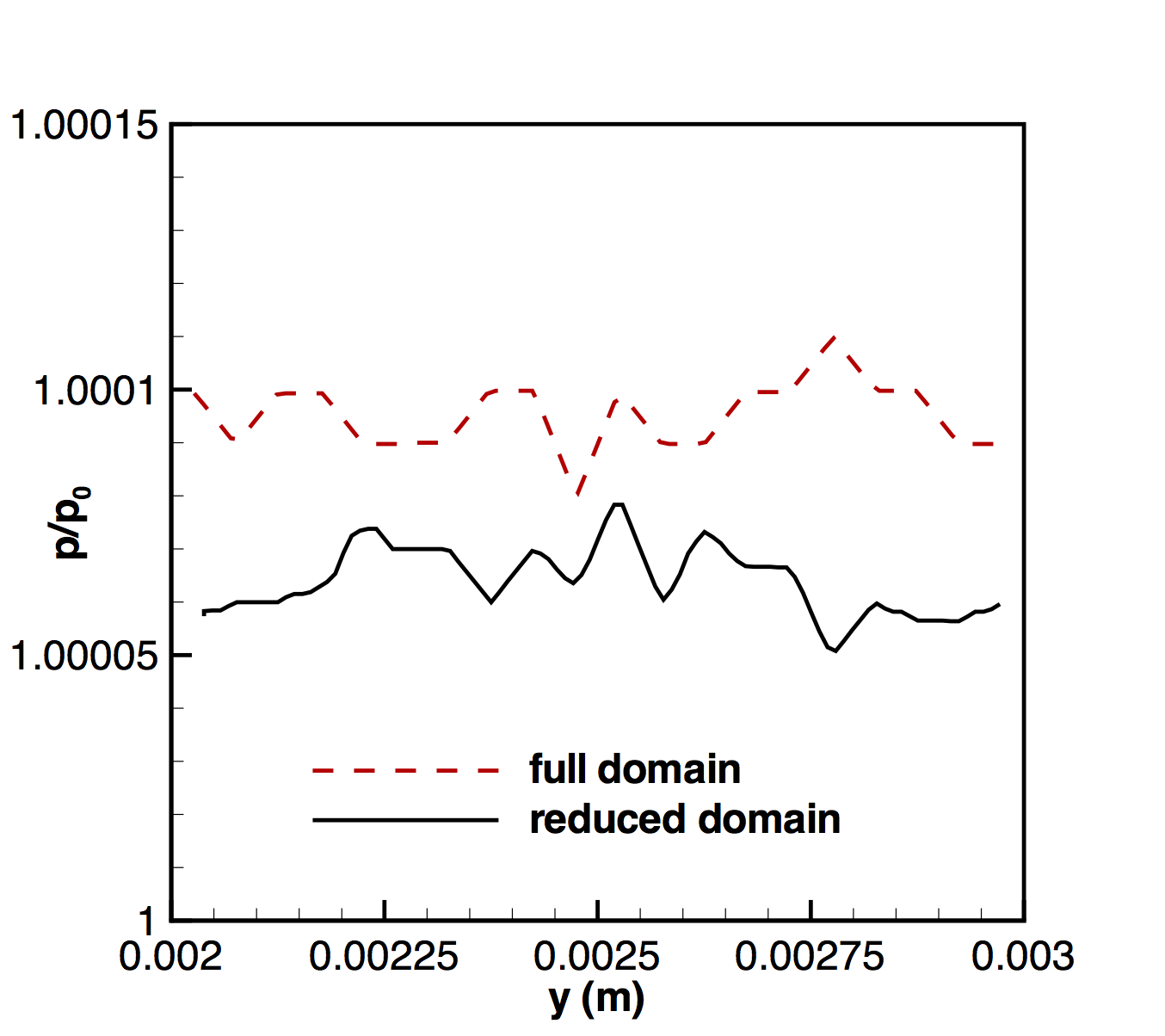}
	\caption{Velocity component $u$ (left) and pressure (right) across the middle station of the micro-channel with $x\equiv L_{\rm all}/2$ and $z\equiv H_{\rm all}/2$.}
	\label{Fig:vectup}
\end{figure}

Fig.~\ref{Fig:vectup} shows the velocity $u$ (left) and pressure (right) profiles at the middle station of the channel extracted from Fig.~\ref{Comparison between ib11 and notanks on XOY}. The velocity of full domain simulation is noticeably smaller than that of the reduced domain simulation, which is consistent with Fig.~\ref{Fig:flatup} (left).  The profiles are parabolic and the velocity slips along the channel surface are evident. The two pressure profiles do not have large fluctuations and they are quite flat with maximum relative fluctuations of $0.01 \%$.  

\section{Mass flow rates of different gas species through real micro-channel at different $Kn_0$}\label{s:Real cases}
Gas flows inside a real micro-channel with $L_{\rm micro}\times W_{\rm micro}\times H_{\rm micro}\equiv73$ mm $\times 6$ mm $\times 0.22$ mm \cite{Yamaguchi2016} are simulated by using $L_{\rm all}\times W_{\rm all}\times H_{\rm all}\equiv80$ mm $\times 10$ mm $\times 1$ mm with $\phi=0.20795$. $\Delta x=\Delta y\gg\Delta z$ are chosen at low pressure conditions to optimize the cell division. The three cell sizes are always smaller than $\lambda_0$ at different pressure conditions as required, e.g., a total of $3200\times400\times50$ cells are used for argon gas flow at $\delta_0(p_0=294 {\rm Pa})=7.41$ with $\lambda_0\approx0.0268$ mm (\textit{note}: this simulation takes about one day for 2000 time steps when using 40 CPU cores), where $\delta_0$ is a mean rarefaction parameter to characterize the mass flow rate $\dot{M}$ \cite{Yamaguchi2016}: 
\begin{equation}\label{eq:delta}
\begin{aligned}
   \delta_0=\dfrac{p_0H_{\rm micro}}{\mu_0\sqrt{2k_{\rm B}T_0/m}}\approx\dfrac{0.9025}{Kn_0}. 
\end{aligned}
\end{equation}

In addition to argon, we use $\mu=1.865\times10^{-5}\times(T/273)^{0.66} {\rm Pa\cdot s}$ for helium molecules ($m=6.65\times10^{-27}$ kg) and $\mu=2.975\times10^{-5}\times(T/273)^{0.66} {\rm Pa\cdot s}$ for neon molecules ($m=33.5\times10^{-27}$ kg) \cite{Shen2005}. Pure Maxwell diffuse reflection model is used at the wall surface as in the previous tests. Since the relative density variation inside the micro-channel is very small, the volumetric velocity component $\left<u\right>_{\Omega}$ in the $x$ direction at steady state is used to compute $\dot{M}_{\rm DSBGK}$ as follows: 
\begin{equation}\label{eq:flow rate}
\begin{aligned}
   \dot{M}_{\rm DSBGK}&=\left<u\right>_\Omega mn_0W_{\rm micro}H_{\rm micro} \\&=\dfrac{(\sum_jn_jV_ju_j)mn_0W_{\rm micro}H_{\rm micro}}{n_0L_{\rm micro}W_{\rm micro}H_{\rm micro}} \\&=\dfrac{m\sum_jn_jV_ju_j}{L_{\rm micro}},
\end{aligned}
\end{equation}
where $n_j$, $V_j\equiv\Delta x\Delta y\Delta z$ and $u_j$ are the number density, volume and flow velocity component of the cell $j$ inside the micro-channel, respectively, and $n_0=p_0/(k_{\rm B}T_0)$. Similarly, the average velocity components $\left<u\right>_{\partial\Omega,{\rm in}}$ and $\left<u\right>_{\partial\Omega,{\rm out}}$ are also computed by using the summations over cells on the sections at the inlet and outlet of the micro-channel, respectively. The purpose is to monitor local convergence. 

\begin{figure}[H]
	\centering
	\includegraphics[width=0.6\textwidth]{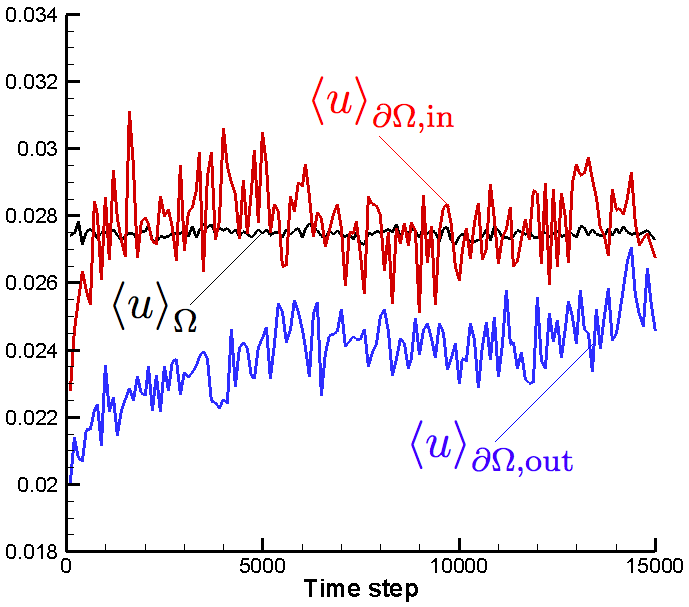}
	\caption{Convergence processes of average velocity components in the simulation of argon flow, $T_{\rm H}=347.1$ K, $T_{\rm L}=289.2$ K and $\delta_0=1.70$.}
	\label{convergence of mean u}
\end{figure}

\begin{figure}[H]
	\centering
	\includegraphics[width=0.6\textwidth]{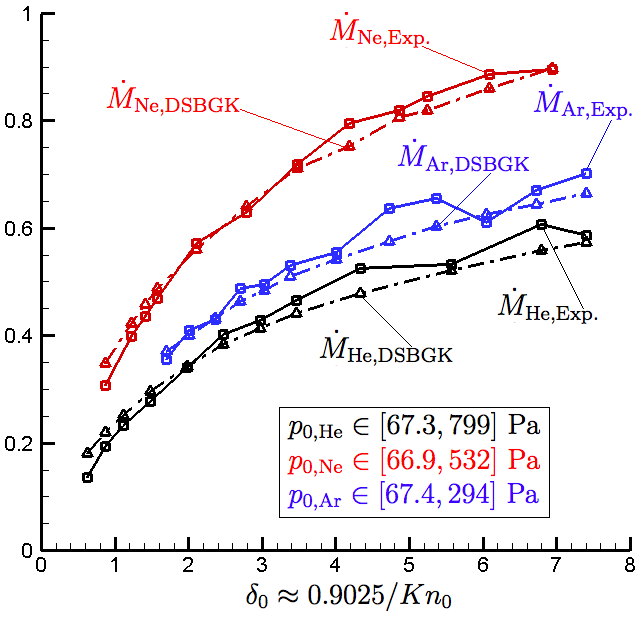}
	\caption{Mass flow rates $\dot{M}$ ($\times10^{-10}$ kg/s) computed by the DSBGK method and the experimental data \cite{Yamaguchi2016}, $T_{\rm H}=347.1$ K and $T_{\rm L}=289.2$ K.}
	\label{compare flow rate T1}
\end{figure}

The convergence processes of three average velocity components in a representative case are given in Fig.~\ref{convergence of mean u}, which shows that $\left<u\right>_\Omega$ converges much faster than the local quantities. Thus, the computational cost can be reduced in studying the mass flow rate by using the global quantity $\left<u\right>_\Omega$. Fig.~\ref{compare flow rate T1} and Table~\ref{tab:compare flow rate T1} show that the DSBGK results agree very well with the experimental data over a wide rage of $\delta_0$ for different gas species. Fig.~\ref{compare flow rate T1} also shows that the DSBGK results have smoother and milder variations with $\delta_0$ than the experimental results. 

\newsavebox{\tableboxa} 
\begin{lrbox}{\tableboxa} 
\begin{tabular}{c c c | c c c | c c c}
\hline
\multicolumn{3}{c|}{{\bf{He}}, $p_0\in[67.3, 799]$ Pa} & \multicolumn{3}{c|}{{\bf{Ne}}, $p_0\in[66.9, 532]$ Pa} & \multicolumn{3}{c}{{\bf{Ar}}, $p_0\in[67.4, 294]$ Pa}\\
$\delta_0$ & $\dot{M}_{\rm Exp.}$ & $\dot{M}_{\rm DSBGK}$ & $\delta_0$ & $\dot{M}_{\rm Exp.}$ & $\dot{M}_{\rm DSBGK}$ & $\delta_0$ & $\dot{M}_{\rm Exp.}$ & $\dot{M}_{\rm DSBGK}$ \\
\hline
.624  & 0.137 & 0.180 & .873 & 0.307 & 0.349 & 1.70 & 0.355 & 0.371 \\
.865  & 0.193 & 0.219 & 1.22 & 0.399 & 0.423 & 2.01 & 0.409 & 0.401 \\
1.11  & 0.233 & 0.252 & 1.41 & 0.435 & 0.459 & 2.36 & 0.430 & 0.432 \\
1.48  & 0.278 & 0.296 & 1.58 & 0.469 & 0.488 & 2.70 & 0.487 & 0.464 \\
1.98  & 0.341 & 0.342 & 2.10 & 0.571 & 0.560 & 3.03 & 0.496 & 0.485 \\
2.47  & 0.402 & 0.383 & 2.78 & 0.629 & 0.640 & 3.38 & 0.531 & 0.511 \\
2.98  & 0.429 & 0.414 & 3.48 & 0.718 & 0.711 & 4.01 & 0.554 & 0.542 \\
3.47  & 0.465 & 0.441 & 4.18 & 0.795 & 0.753 & 4.73 & 0.636 & 0.575 \\
4.33  & 0.526 & 0.479 & 4.87 & 0.819 & 0.806 & 5.37 & 0.655 & 0.604 \\
5.57  & 0.532 & 0.521 & 5.24 & 0.845 & 0.819 & 6.04 & 0.611 & 0.625 \\
6.80  & 0.607 & 0.559 & 6.09 & 0.886 & 0.861 & 6.72 & 0.670 & 0.645 \\
7.41  & 0.587 & 0.574 & 6.94 & 0.896 & 0.897 & 7.41 & 0.702 & 0.664 \\
\hline
\end{tabular}
\end{lrbox}

\begin{table*}[!htb] 
\caption{Mass flow rates $\dot{M}$ ($\times10^{-10}$ kg/s) computed by the DSBGK method and the experimental data \cite{Yamaguchi2016}, $T_{\rm H}=347.1$ K and $T_{\rm L}=289.2$ K.}\label{tab:compare flow rate T1}
\begin{center}
\resizebox{0.99\textwidth}{!}{\usebox{\tableboxa}} 
\end{center}
\end{table*}




%
\section{Further comparison in a 2D case}\label{s:2D case}
\begin{figure}[H]
	\centering
	\includegraphics[width=0.44\textwidth]{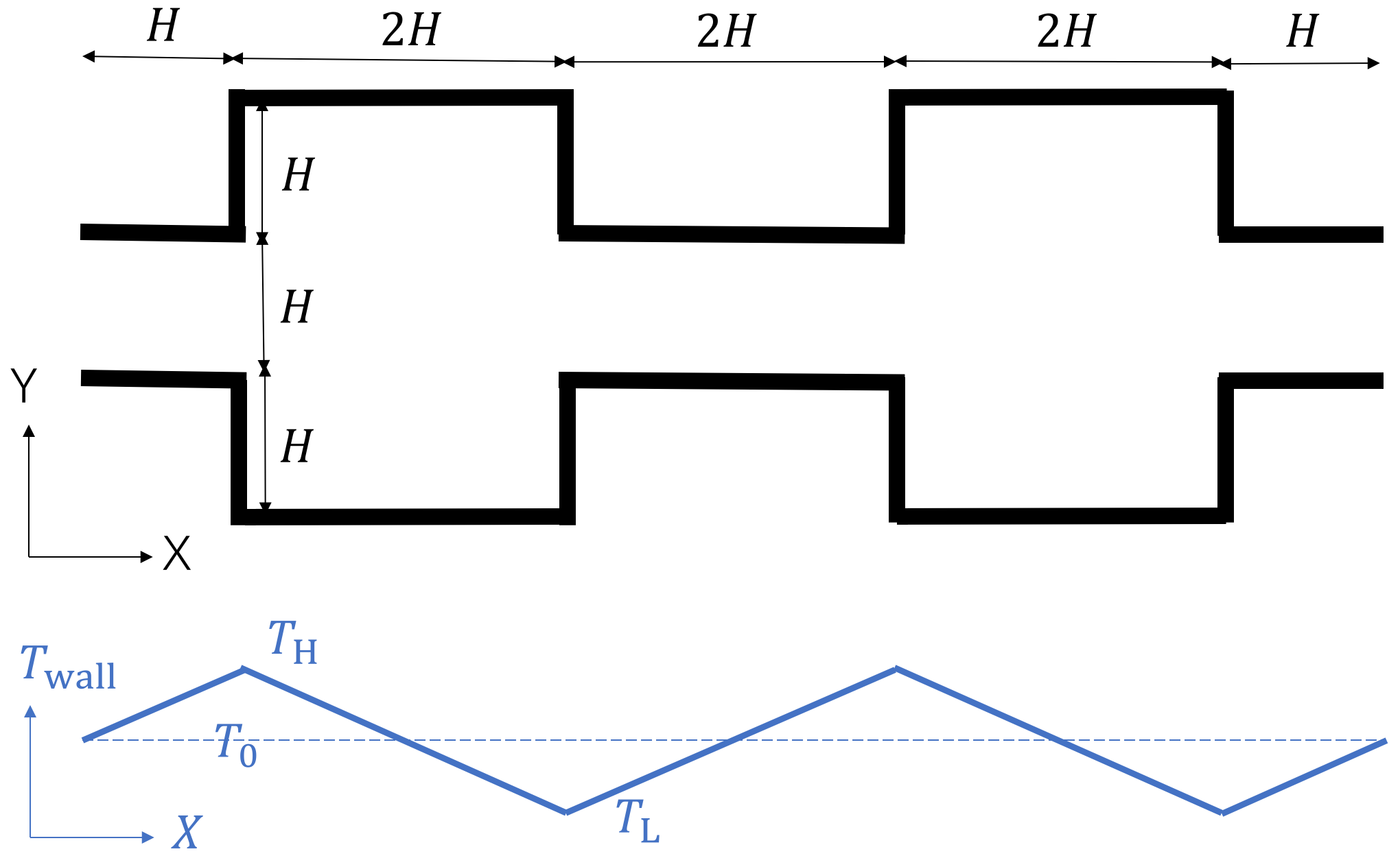}
	\caption{Thermal transpiration gas flow inside a 2D channel.}
	\label{2D geometry}
\end{figure}

\begin{figure}[H]
	\centering
	\includegraphics[width=0.35\textwidth]{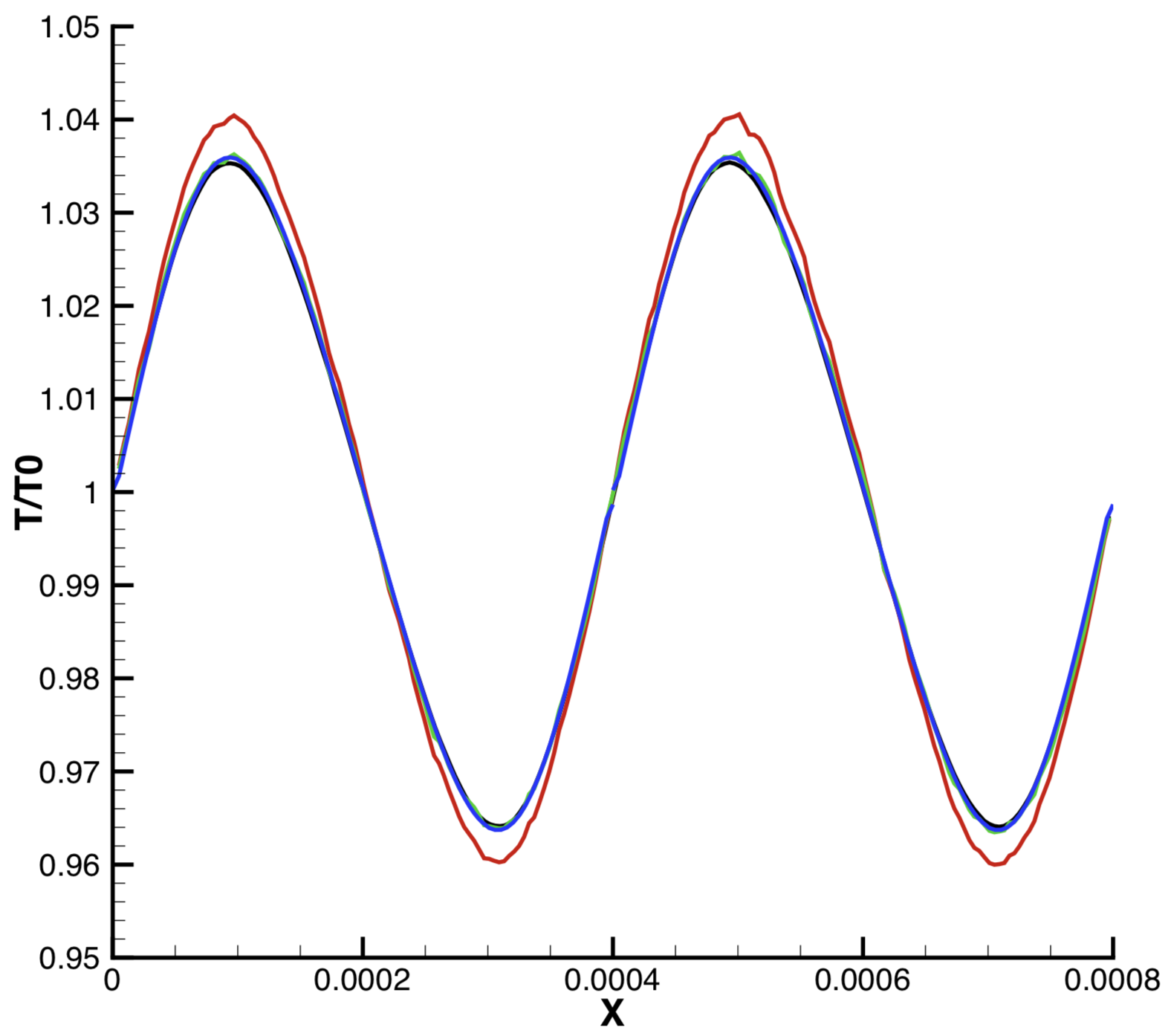}
	\includegraphics[width=0.35\textwidth]{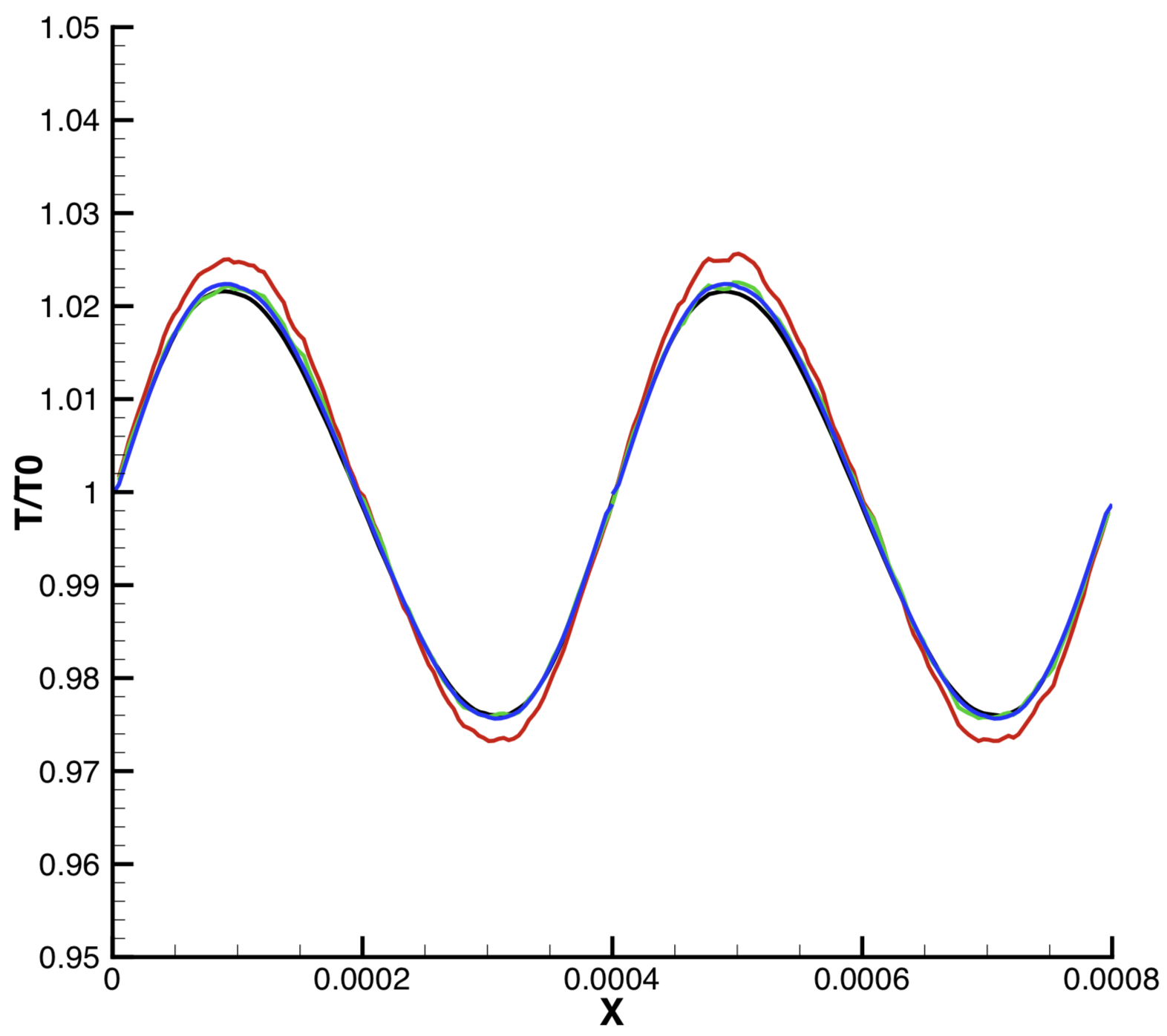}\\	
	\includegraphics[width=0.35\textwidth]{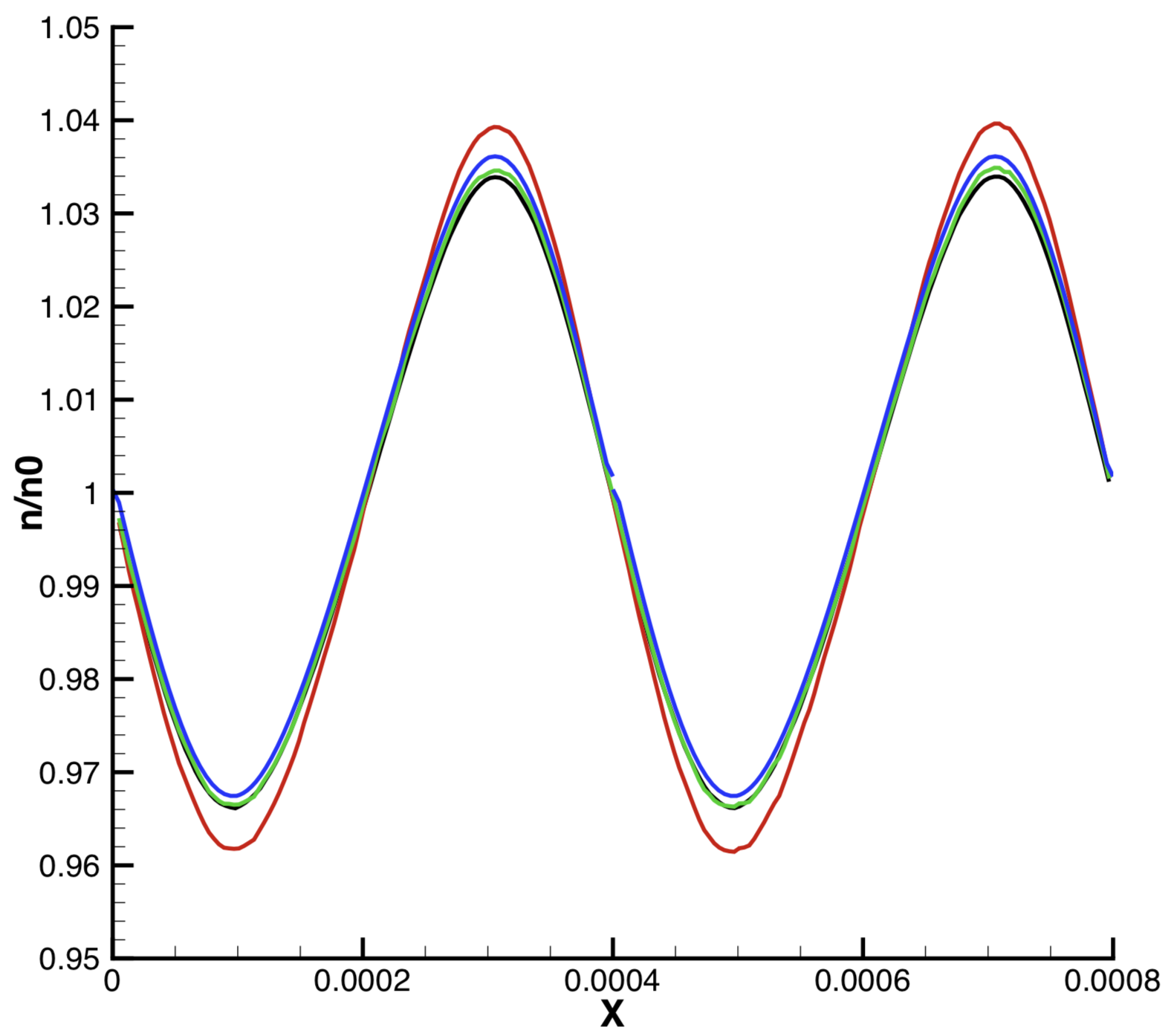}
	\includegraphics[width=0.35\textwidth]{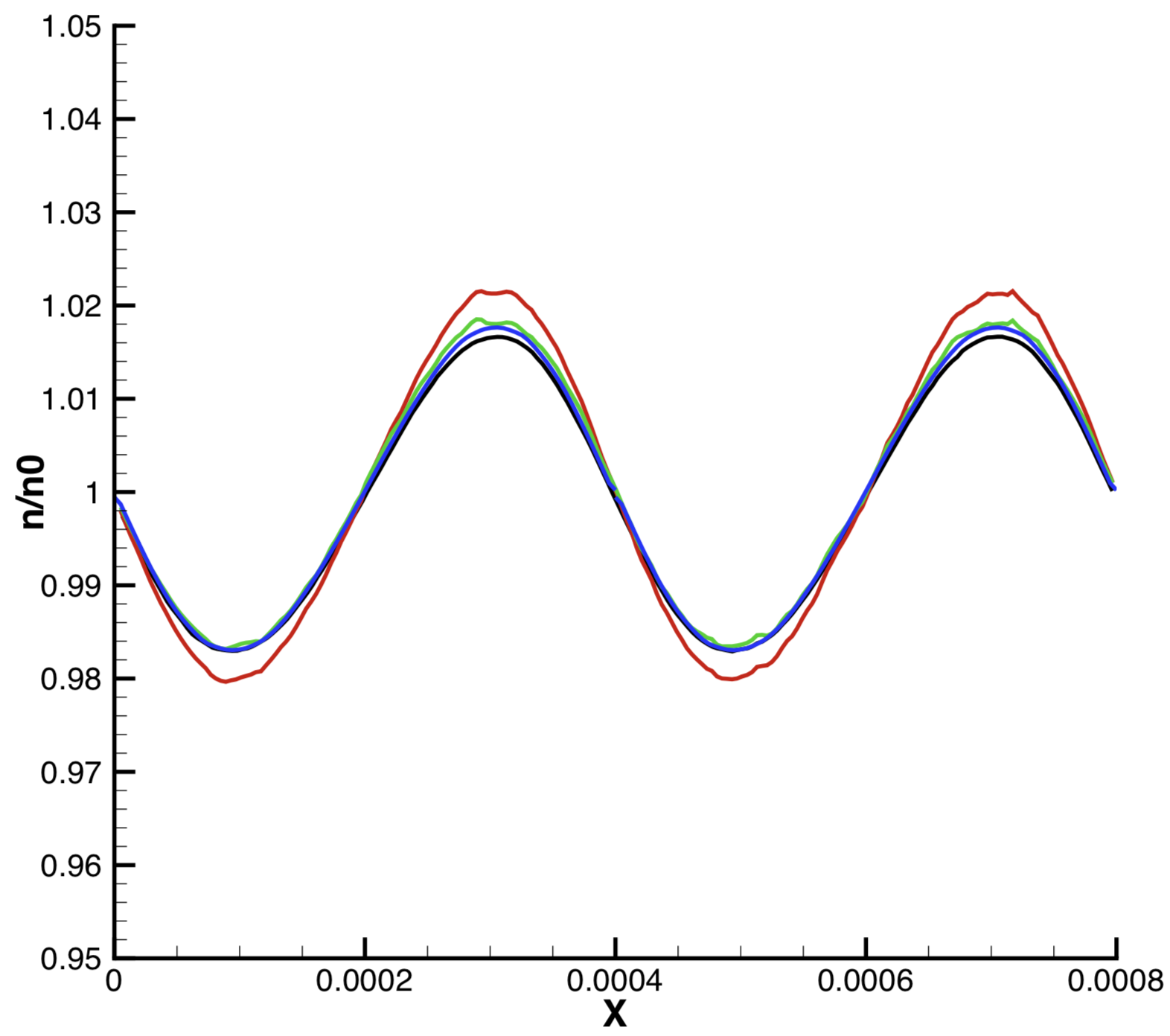}\\	
	\includegraphics[width=0.35\textwidth]{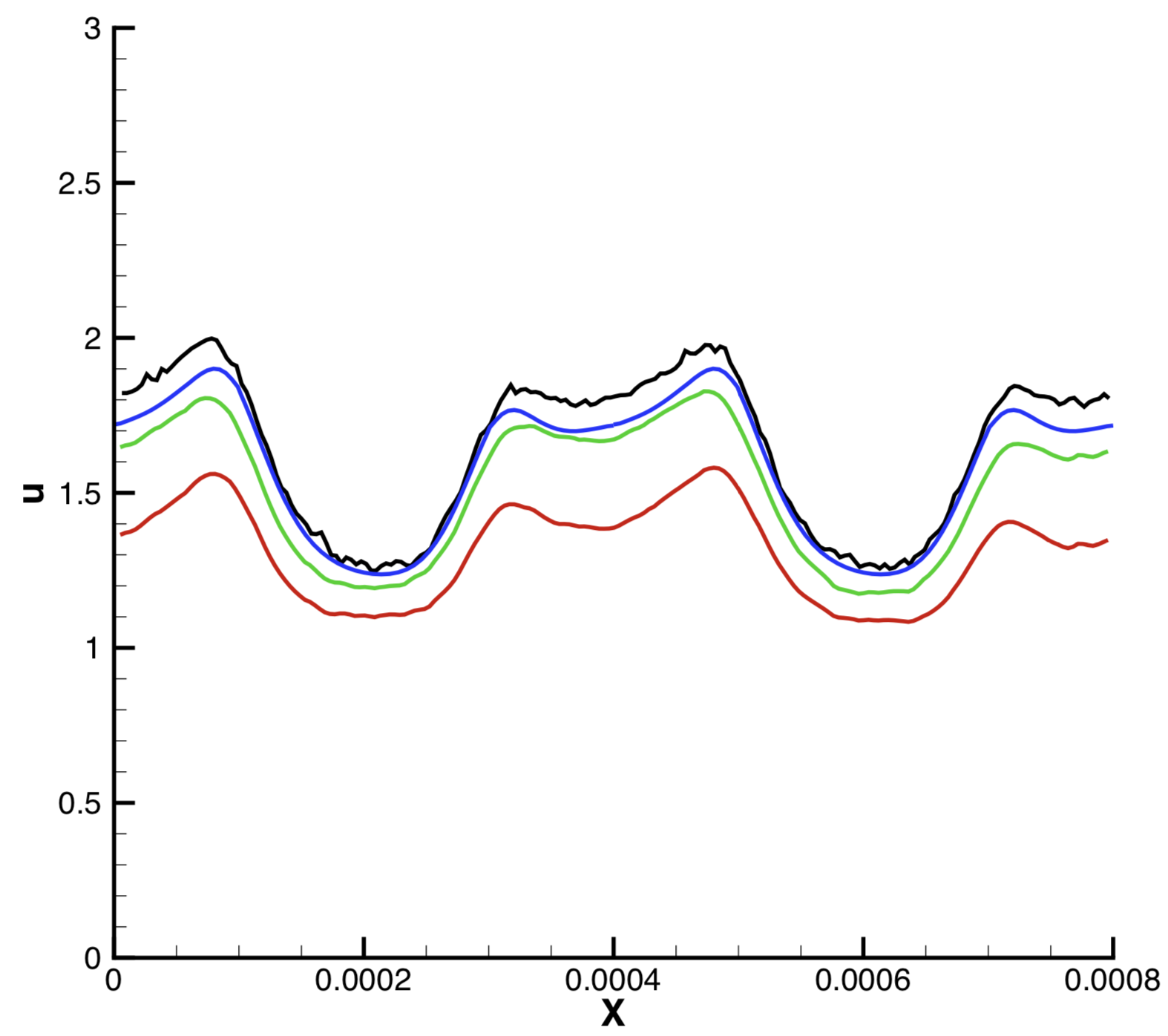}
	\includegraphics[width=0.35\textwidth]{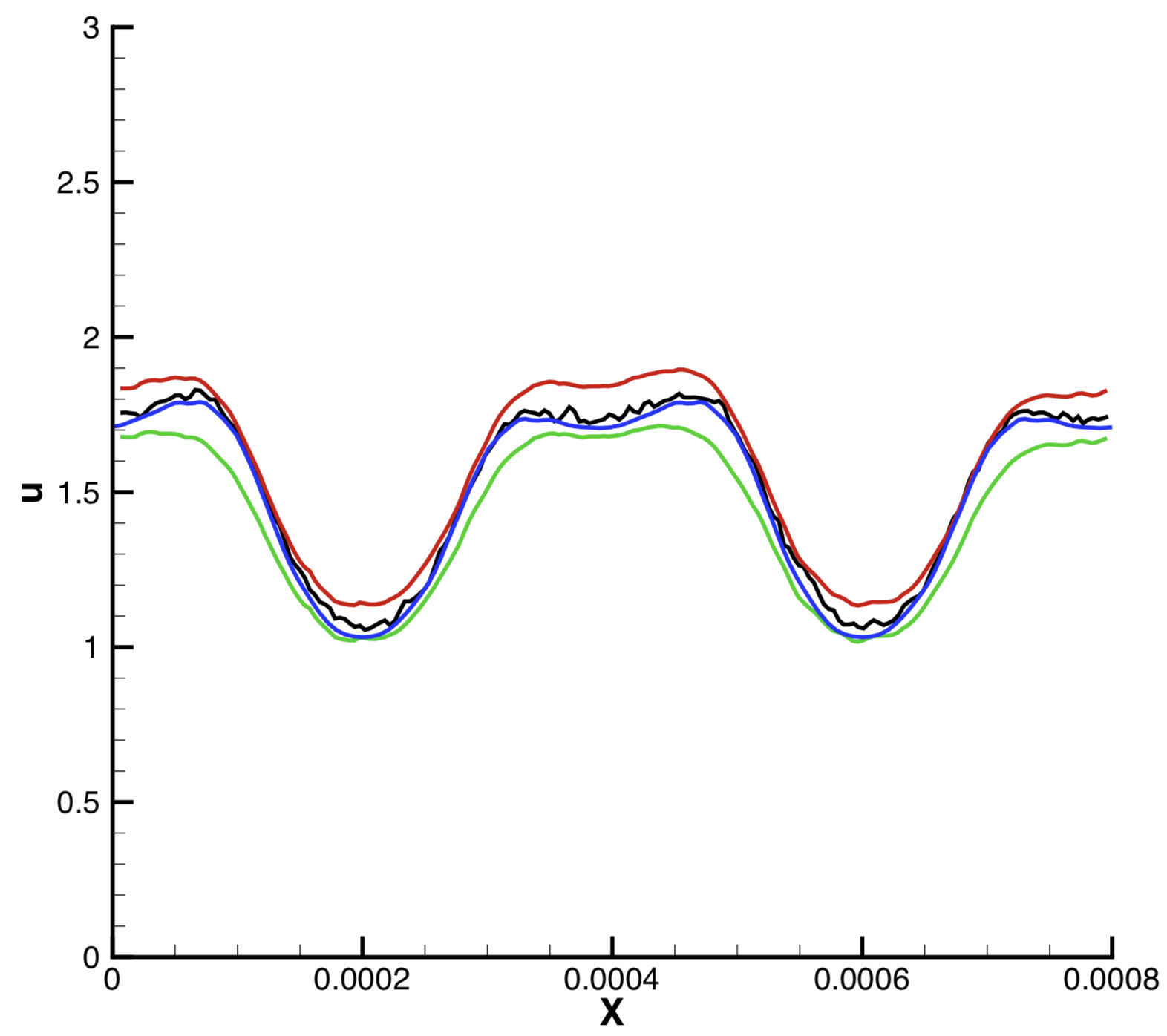}\\	
	\caption{Comparison between different numerical methods.}
	\label{Line-DSBGKvsDSMC}
\end{figure}

The schematic of a 2D thermal transpiration argon gas flow is given in Fig.~\ref{2D geometry} and  Fig.~\ref{Line-DSBGKvsDSMC} shows the comparison on the centerline between the DSMC result (black), the solution of Shakhov model equation (blue), and the DSBGK result of matching heat conductivity coefficient via $\upsilon=2nk_{\rm B}T/(3\mu)$ (green), at $Kn_0=0.2$ (left) and $Kn_0=1$ (right), respectively. We also present the DSBGK result of matching viscosity via $\upsilon=nk_{\rm B}T/\mu$  (red) as the ordinary implementation of BGK equation that has noticeable error in the thermal transpiration flow problems as also reported elsewhere. 

\section{Conclusions}\label{s:Conclusions}
Numerical simulations of  thermal transpiration flows through a micro-channel with different species at different $Kn$ are performed with the DSBGK method. Simulation setup effects on the final simulation results are discussed.

It is found that for flows of several species with different degrees of rarefaction, the mass flow rates predicted by the simulations and  measurements agree quite well.  It indicates that the DSBGK method is more superior than traditional particle simulation methods that are subject to large statistical noises in simulating low-speed micro gas flows. Meanwhile, the DSBGK method can simulate micro flow with high rarefaction, which is a serious challenge to traditional computation schemes based on the continuum flow assumption. 

This study also indicates that simulation cost without reservoirs can be much lower than that of full domain simulation. However, their final flow field patterns are different because the confinement effect (inhaling and exhaling) is neglected in the reduced domain simulation without reservoirs. The confinement effect happens in outside regions close to the micro-channel ends and changes the inlet and outlet conditions of the micro-channel. It is easy to understand that simulations with reservoirs attached to the micro-channel ends are closer to the real experiments, and thus the results are more accurate. Simulations without reservoirs could develop much faster but the difference between the mass flow rates computed with reservoirs and without reservoirs is appreciable. Thus, it is questionable to determine the surface momentum accommodation coefficient by using analytical solution of the mass flow rate obtained in a single-channel problem without the confinement effect of reservoirs at the two ends. 

\section{Acknowledgements}\label{s:Acknowledge}
J. Li thanks Prof. Irina Graur for her helpful suggestions.  




\begin{thebibliography}{17}
\expandafter\ifx\csname
natexlab\endcsname\relax\def\natexlab#1{#1}\fi
\expandafter\ifx\csname bibnamefont\endcsname\relax
  \def\bibnamefont#1{#1}\fi
\expandafter\ifx\csname bibfnamefont\endcsname\relax
  \def\bibfnamefont#1{#1}\fi
\expandafter\ifx\csname citenamefont\endcsname\relax
  \def\citenamefont#1{#1}\fi
\expandafter\ifx\csname url\endcsname\relax
  \def\url#1{\texttt{#1}}\fi
\expandafter\ifx\csname urlprefix\endcsname\relax\def\urlprefix{URL
}\fi \providecommand{\bibinfo}[2]{#2}
\providecommand{\eprint}[2][]{\url{#2}}

\bibitem{Reynolds}
O. Reynods, ``On certain dimensional properties of matter in the gaseous temperature,'' {\color{blue} Philos. Trans. R. Soc. London}, {\textbf{170}}, 727-845 (1879).

\bibitem{Maxwell}
J. Maxwell, ``On stresses in rarefied gases arising from inequilities of temperature,''  {\color{blue} Philos. Trans. R. Soc. London}, {\textbf{170}}, 231-256 (1879).

\bibitem{Knudsen}
M. Knudsen, ``Eine revision der gleichgewichtsbedingung der gase. Thermische molekularstromung,'' {\color{blue} Ann. Phys.}, {\textbf{336}}, 205-229 (1909).

\bibitem{Shen2005}
C. Shen,  {\textit{Rarefied Gas Dynamics: Fundamentals, Simulations and Micro Flows}}, Springer. (2005).

\bibitem{Ali}
G. Karniadakis, A. Beskok and N. Alura, {\it Microflows and Nanoflows: Fundamentals and Simulation}, 2005,
Springer-Verlag, New York. ISBN: 978-0-387-22197-7. doi:10.1007/0-387-28676-4.

\bibitem{Vargo}
S. Vargo, E. Muntz, G. Shiflett and W.Tang, ``Kunden compressor as a micro- and macro- scale vacuum pump without moving parts or fluids,'' {\color{blue} J. Vac. Sci. Technol. A}, {\textbf{17}}, 2308 (1999).

\bibitem{Young}
M. Young, Y. Han, E. Muntz, G. Shiflett, A. Ketsdever and A. Green, ``Theraml transpiration in micro-sphere membrances,'' {\color{blue}AIP Conf. Proc.}, {\textbf{663}}, 743-751 (2003).

\bibitem{Alexeenko}
A. Alexeenko, S. Gimelshein, E. Muntz, and A. Ketsdever, ``Kinetic modeling of temperature driven flows in short microchannels,'' {\color{blue} Int. J. Therm. Sci}, {\textbf{45}}, 1045-1051(2006).

\bibitem{Gupta}
N. K. Gupta, S. An and Y.B. Gianchandani, ``A Si-micromachined 48-stage Knudsen pump for on-chip vacuum,'' {\color{blue}J. Micromech. Microeng.}, {\textbf{22}}, 105026 (2012).

\bibitem{Liang}
S. Liang, ``Some measurements of thermal transpiration,'' {\color{blue} J. Appl. Phys.}, {\textbf{22}}, 148 (1951).

\bibitem{Weber}
S. Weber and G. Schmidt, {\color{blue} Commun. Leiden. Rapp. et Commun}, {\textbf{246c}}, 72 (1936).

\bibitem{Rosenberg}
A. Rosenberg and C. Martel Jr., ``Theraml transpiration of gases at low pressures,'' {\color{blue} J. Phys. Chem}, {\textbf{62}}, 457-459 (1958).

\bibitem{marcros1}
M.R. Cardenas, I. Graur, P. Perrier and J.G. Meolans, ``Time-dependent experimental analysis of a thermal transpiration rarefied gas flow,'' {\color{blue}Phys. Fluids}, {\textbf{25}}, 072001 (2013). doi:10.1063/1.4813805.

\bibitem{marcros2}
M.R. Cardenas, I. Graur, P. Perrier and J.G. Meolans, ``Thermal transpiration flow: a circular cross-section microtube submitted to a temperature gradients,'' {\color{blue}Phys. Fluids}, {\textbf{23}}, 031702 (2011).

\bibitem{marcros3}
M.R. Cardenas, I. Graur, P. Perrier and J. Meolans, ``An experimental and numerical study of the final zero-flow theraml transpiration stage,'' {\color{blue}J. Therm. Sci. Technol}, {\textbf{7}}, 437-452 (2012).

\bibitem{los}
J. Los and R. Fergusson, ``Measurements of thermomolecular pressure differences on argon and nitrogen,'' {\color{blue}Trans. Faraday Soc.}, {\textbf{48}}, 730-738 (1952).

\bibitem{Takaishi}
T. Takaishi and Y. Sensui, ``Thermal  transpiration effect of hydrogen, rare gases and methane,'' {\color{blue} Trans. Faraday Soc.}, {\textbf{59}}, 2503-2514 (1963).

\bibitem{annis}
B. Annis, ``Thermal creep in gases,'' {\color{blue} J. Chem. Phys.}, {\textbf{57}}, 2898 (1972).

\bibitem{Loyalka}
S. Loyalka and J. Cipolla Jr., ``Thermal creep slip with arbitary accommondation at the surface,'' {\color{blue}Phys. Fluids}, {\textbf{14}}, 1656 (1971).

\bibitem{sone}
Y. Sone and H. Sugimoto, ``Vacuum pump without a moving part and its performance,'' {\color{blue} AIP Conf. Proc.}, {\textbf{663}}, 1041 (2003).

\bibitem{Sugimoto}
H. Sugimoto, S. Kawakami and K. Moriuchi, ``Rarefied gas flows induced through a pair of parallel  meshes with different temperatures,'' {\color{blue} AIP Conf. Porc.}, {\textbf{1084}}, 1021 (2008).

\bibitem{Ewart}
T. Ewart, P. Perrier, I. Graur and J. G. Meolans, ``Mass flow rate measurements in gas micro flows,'' {\color{blue} Exp. Fluids}, {\textbf{41}}, 487-498 (2006).

\bibitem{Yamaguchi2014}
H. Yamaguchi, M.R. Cardenas, P. Perrier, I. Graur and T. Niimi, ``Thermal transpiration flow through a single rectangular channel,'' {\color{blue} J. Fluid Mech.}, {\textbf{744}}, 169-182 (2014).

\bibitem{Yamaguchi2016}
H. Yamaguchi, P. Perrier,  M.T. Ho, J.G. Meolans, T. Niimi and I. Graur, ``Mass flow  rate measurement of thermal creep flow from transitional to slip flow regime,'' {\color{blue}{J. Fluid Mech.}}, {\textbf{795}}, 690-707 (2016).

\bibitem{BGK}
P.L. Bhatnagar, E.P. Gross and M. Krook, ``A model for colliion processes in gases I: small amplitude processes in charged and neutral one-component systems,'' {\color{blue} Phys. Rev.}, {\textbf{94}}, 511-525 (1954).

\bibitem{Qian1992}
Y.H. Qian, D. d'Humieres and P. Lallemand, "Lattice BGK models for Navier-Stokes equation," {\color{blue}{Europhysics Letters}}, {\textbf{17}}, 479-484 (1992). 

\bibitem{luo}
X. He and L. Luo,  ``Theory of the lattice Boltzmann method: from the Boltzmann equation to the lattice Boltzmann equation,'' {\color{blue} Phys. Rev. E}, {\textbf{56}}, 6811 (1997).

\bibitem{chen}
S. Chen and G. Doolen, ``Lattice Boltzmann method for fluid flows,''  {\color{blue} Annual Rev. Fluid Mech.}, {\textbf{30}}, 329-364 (1998).  https://doi.org/10.1146/annurev.fluid.30.1.329.

\bibitem{Sharipov}
I. Graur and F. Sharipov, ``Non-isothermal flow of rarefied gas through a long piple with elliptic cross section,'' {\color{blue}Microfluid. Nonofluid.}, {\textbf{6}}, 267-275 (2009).

\bibitem{xugks}
K. Xu, ``A gas-kinetic BGK scheme for the Navier-Stokes equations and its connection with artificial dissipation and Godunov method,'' {\color{blue} J. Comput. Phys.}, {\textbf{171}},  289–335 (2001).  doi:10.1006/jcph.2001.6790.

\bibitem{xu}
K. Xu and Z.H. Li, ``Microchannel flow in the slip regime: gas-kinetic BGK-Burnett solutions,''  {\color{blue}J. Fluid. Mech.}, {\textbf{513}},  87-110 (2004). https://doi.org/10.1017/S0022112004009826.

\bibitem{xuuks}
K. Xu and J.C. Huang, ``A unified gas kinetic scheme for continuum and rarefied flows,'' {\color{blue}J. Comput. Phys.}, {\textbf{229}} (2010).

\bibitem{Bird1963}
G. A. Bird, ``Approach to translational equilibrium in a rigid sphere gas,'' {\color{blue}{Phys. Fluids}}, {\textbf{6}}, 1518 (1963).

\bibitem{Bird1994}
G. A. Bird, {\textit{Molecular Gas Dynamics and the Direct Simulation of Gas Flows}}, Clarendon Press, Oxford (1994).

\bibitem{fan0}
J. Fan  and C. Shen, ``Statistical simulation of low-speed unidirectional flows in transition regime'', in {\it Rarefied Gas Dynamics}, edited by, R. Brun, et al., Cepadus-Editions, Toulouse, \textbf{245} (1999).

\bibitem{cai}
C. Cai, I.D. Boyd, J. Fan and G.V. Candler,  ``Direct simulation methods for low-speed microchannel flows,'' {\color{blue} J. Thermophys. Heat Transfer}, {\textbf{14}} (3), 368-378 (2000).

\bibitem{fan1}
J. Fan and  C. Shen, ``Statistical simulation of low speed rarefied gas flows,'' {\color{blue} J. Comput. Phys.}, {\textbf{167}} (2), 393-412 (2001)
	
\bibitem{sun}
Q. Sun and I.D. Boyd, ``A direct simulation method for subsonic, microscale gas flows,'' {\color{blue} J. Comput. Phys.}, {\textbf{179}} (2), 400-425 (2002). https://doi.org/10.1006/jcph.2002.7061.

\bibitem{fan2}
C. Shen,  J. Fan and C. Xie, ``Statistical simulaiton of rarefied gas flows in micro-channles,'' {\color{blue} J. Compt. Phys.}, {\textbf{189}}, 512-526 (2003). https://doi.org/10.1016/S0021-9991(03)00231-6

\bibitem{Li2010}
J. Li, ``Direct simulation method based on BGK equation,'' In \textit{27th International Symposium on Rarefied Gas Dynamics}, AIP Conference Proceedings, \textbf{1333}: 283-288 (2011).

\bibitem{Li2012}
J. Li, ``Comparison between the DSMC and DSBGK methods,'' {\textit{arXiv}: 1207.1040 [physics.comp-ph]} (2012).

\bibitem{LiSultan2015}
J. Li  and A.S. Sultan, ``Permeability computations of shale gas by the pore-scale Monte Carlo molecular simulations,'' In {\textit{International Petroleum Technology Conference}}, IPTC-18263-MS (2015).

\bibitem{LiSultan2016}
J. Li  and A.S. Sultan, ``Klinkenberg slippage effect in the permeability computations of shale gas by the pore-scale simulations,'' {\color{blue}{J. Natural Gas Sci. Engineering}}, in press (2016). https://doi.org/10.1016/j.jngse.2016.07.041.

\bibitem{entrance1}
E.M. Sparrow, S.H. Lin and T.S. Lundgren, ``Flow development in the hydrodynamic entrance region of tubes and ducts,'' {\color{blue} Phys. Fluids}, {\textbf{7}} (3), 338-347 (1964).

\bibitem{entrance2}
Z. Duan and Y. Muzychka, ``Slip flow in the hydrodynamic entrance region of circular and noncircular microchannels,'' {\color{blue} J. Fluids Eng.}, {\textbf{132}} (1), 011201 (2009).

\bibitem{coeff}
E. B. Arklic, K. S. Breuer and M.A. Schmidt, ``Mass flow and tangential momentum accommodation in sillicon micromachined channels,'' {\color{blue}J. Fluid Mech.}, {\textbf{437}}, 29-43 (2001).


\end{thebibliography}
\end{document}